\documentclass[superscriptaddress,twocolumn,secnumarabic,amssymb, nofootinbib, nobibnotes, aps, prd,floatfix]{revtex4-2}

\usepackage{graphicx}
\usepackage{multirow}
\usepackage{orcidlink}
\usepackage{amsmath}

\usepackage{lineno}

\begin{document}
\raggedbottom

\title{Identification of low-energy kaons in the ProtoDUNE-SP detector}%

%

\newcommand{\Albanysuny}{University of Albany, SUNY, Albany, NY 12222, USA}
\newcommand{\Almaty}{Institute of Nuclear Physics at Almaty, Almaty 050032, Kazakhstan
}
\newcommand{\Amsterdam}{University of Amsterdam, NL-1098 XG Amsterdam, The Netherlands}
\newcommand{\Antalya}{Antalya Bilim University, 07190 D\"o{\c s}emealtı/Antalya, Turkey}
\newcommand{\Antananarivo}{University of Antananarivo, Antananarivo 101, Madagascar}
\newcommand{\Antioquia}{University of Antioquia, Medell\'in, Colombia}
\newcommand{\AntonioNarino}{Universidad Antonio Nari\~no, Bogot\'a, Colombia}
\newcommand{\Argonne}{Argonne National Laboratory, Argonne, IL 60439, USA}
\newcommand{\Arizona}{University of Arizona, Tucson, AZ 85721, USA}
\newcommand{\Asuncion}{Universidad Nacional de Asunci\'on, San Lorenzo, Paraguay}
\newcommand{\Athens}{University of Athens, Zografou GR 157 84, Greece}
\newcommand{\Atlantico}{Universidad del Atl\'antico, Barranquilla, Atl\'antico, Colombia}
\newcommand{\Augustana}{Augustana University, Sioux Falls, SD 57197, USA}
\newcommand{\Bern}{University of Bern, CH-3012 Bern, Switzerland}
\newcommand{\Beykent}{Beykent University, Istanbul, Turkey}
\newcommand{\Birmingham}{University of Birmingham, Birmingham B15 2TT, United Kingdom}
\newcommand{\BolognaUniversity}{Universit\`a di Bologna, 40127 Bologna, Italy}
\newcommand{\Boston}{Boston University, Boston, MA 02215, USA}
\newcommand{\Bristol}{University of Bristol, Bristol BS8 1TL, United Kingdom}
\newcommand{\Brookhaven}{Brookhaven National Laboratory, Upton, NY 11973, USA}
\newcommand{\Bucharest}{University of Bucharest, Bucharest, Romania}
\newcommand{\CalBerkeley}{University of California Berkeley, Berkeley, CA 94720, USA}
\newcommand{\CalDavis}{University of California Davis, Davis, CA 95616, USA}
\newcommand{\CalIrvine}{University of California Irvine, Irvine, CA 92697, USA}
\newcommand{\CalLosangeles}{University of California Los Angeles, Los Angeles, CA 90095, USA}
\newcommand{\CalRiverside}{University of California Riverside, Riverside CA 92521, USA}
\newcommand{\CalSantabarbara}{University of California Santa Barbara, Santa Barbara, CA 93106, USA}
\newcommand{\Caltech}{California Institute of Technology, Pasadena, CA 91125, USA}
\newcommand{\Cambridge}{University of Cambridge, Cambridge CB3 0HE, United Kingdom}
\newcommand{\Campinas}{Universidade Estadual de Campinas, Campinas - SP, 13083-970, Brazil}
\newcommand{\CataniaUniversitadi}{Universit\`a di Catania, 2 - 95131 Catania, Italy}
\newcommand{\Catolica}{Universidad Cat\'olica del Norte, Antofagasta, Chile}
\newcommand{\CBPF}{Centro Brasileiro de Pesquisas F\'isicas, Rio de Janeiro, RJ 22290-180, Brazil}
\newcommand{\CEASaclay}{IRFU, CEA, Universit\'e Paris-Saclay, F-91191 Gif-sur-Yvette, France}
\newcommand{\CERN}{CERN, The European Organization for Nuclear Research, 1211 Meyrin, Switzerland}
\newcommand{\Charles}{Institute of Particle and Nuclear Physics of the Faculty of Mathematics and Physics of the Charles University, 180 00 Prague 8, Czech Republic }
\newcommand{\Chicago}{University of Chicago, Chicago, IL 60637, USA}
\newcommand{\ChungAng}{Chung-Ang University, Seoul 06974, South Korea}
\newcommand{\CIEMAT}{CIEMAT, Centro de Investigaciones Energ\'eticas, Medioambientales y Tecnol\'ogicas, E-28040 Madrid, Spain}
\newcommand{\Cincinnati}{University of Cincinnati, Cincinnati, OH 45221, USA}
\newcommand{\Cinvestav}{Centro de Investigaci\'on y de Estudios Avanzados del Instituto Polit\'ecnico Nacional (Cinvestav), Mexico City, Mexico}
\newcommand{\Colima}{Universidad de Colima, Colima, Mexico}
\newcommand{\ColoradoBoulder}{University of Colorado Boulder, Boulder, CO 80309, USA}
\newcommand{\ColoradoState}{Colorado State University, Fort Collins, CO 80523, USA}
\newcommand{\Columbia}{Columbia University, New York, NY 10027, USA}
\newcommand{\conida}{Comisi\'on Nacional de Investigaci\'on y Desarrollo Aeroespacial, Lima, Peru}
\newcommand{\Cti}{Centro de Tecnologia da Informacao Renato Archer, Amarais - Campinas, SP - CEP 13069-901}
\newcommand{\CUSB}{Central University of South Bihar, Gaya, 824236, India
}
\newcommand{\CzechAcademyofSciences}{Institute of Physics, Czech Academy of Sciences, 182 00 Prague 8, Czech Republic}
\newcommand{\CzechTechnical}{Czech Technical University, 115 19 Prague 1, Czech Republic}
\newcommand{\DannecyleVieux}{Laboratoire d'Annecy de Physique des Particules, Universit\'e Savoie Mont Blanc, CNRS, LAPP-IN2P3, 74000 Annecy, France}
\newcommand{\Daresbury}{Daresbury Laboratory, Cheshire WA4 4AD, United Kingdom}
\newcommand{\Dordt}{Dordt University, Sioux Center, IA 51250, USA}
\newcommand{\Drexel}{Drexel University, Philadelphia, PA 19104, USA}
\newcommand{\Duke}{Duke University, Durham, NC 27708, USA}
\newcommand{\Durham}{Durham University, Durham DH1 3LE, United Kingdom}
\newcommand{\Edinburgh}{University of Edinburgh, Edinburgh EH8 9YL, United Kingdom}
\newcommand{\EIA}{Universidad EIA, Envigado, Antioquia, Colombia}
\newcommand{\Eotvos}{E\"otv\"os Lor\'and University, 1053 Budapest, Hungary}
\newcommand{\erciyes}{Erciyes University, Kayseri, Turkey}
\newcommand{\FCULport}{Faculdade de Ci\^encias da Universidade de Lisboa - FCUL, 1749-016 Lisboa, Portugal}
\newcommand{\FederaldeAlfenas}{Universidade Federal de Alfenas, Po{\c c}os de Caldas - MG, 37715-400, Brazil}
\newcommand{\FederaldeGoias}{Universidade Federal de Goias, Goiania, GO 74690-900, Brazil}
\newcommand{\FederaldoABC}{Universidade Federal do ABC, Santo Andr\'e - SP, 09210-580, Brazil}
\newcommand{\FederaldoRio}{Universidade Federal do Rio de Janeiro, Rio de Janeiro - RJ, 21941-901, Brazil}
\newcommand{\Fermi}{Fermi National Accelerator Laboratory, Batavia, IL 60510, USA}
\newcommand{\Ferrarauniv}{University of Ferrara, Ferrara, Italy}
\newcommand{\Florida}{University of Florida, Gainesville, FL 32611-8440, USA}
\newcommand{\Floridastate}{Florida State University, Tallahassee, FL, 32306 USA}
\newcommand{\Fluminense}{Fluminense Federal University, 9 Icara\'i Niter\'oi - RJ, 24220-900, Brazil }
\newcommand{\Genova}{Universit\`a degli Studi di Genova, Genova, Italy}
\newcommand{\Georgian}{Georgian Technical University, Tbilisi, Georgia}
\newcommand{\Granada}{University of Granada \& CAFPE, 18002 Granada, Spain}
\newcommand{\GranSasso}{Gran Sasso Science Institute, L'Aquila, Italy}
\newcommand{\GranSassoLab}{Laboratori Nazionali del Gran Sasso, L'Aquila AQ, Italy}
\newcommand{\Grenoble}{University Grenoble Alpes, CNRS, Grenoble INP, LPSC-IN2P3, 38000 Grenoble, France}
\newcommand{\Guanajuato}{Universidad de Guanajuato, Guanajuato, C.P. 37000, Mexico}
\newcommand{\Harish}{Harish-Chandra Research Institute, Jhunsi, Allahabad 211 019, India}
\newcommand{\Hawaii}{University of Hawaii, Honolulu, HI 96822, USA}
\newcommand{\hkust}{Hong Kong University of Science and Technology, Kowloon, Hong Kong, China}
\newcommand{\Houston}{University of Houston, Houston, TX 77204, USA}
\newcommand{\Hyderabad}{University of  Hyderabad, Gachibowli, Hyderabad - 500 046, India}
\newcommand{\Idaho}{Idaho State University, Pocatello, ID 83209, USA}
\newcommand{\IFIC}{Instituto de F\'isica Corpuscular, CSIC and Universitat de Val\`encia, 46980 Paterna, Valencia, Spain}
\newcommand{\IGFAE}{Instituto Galego de F\'isica de Altas Enerx\'ias, University of Santiago de Compostela, Santiago de Compostela, 15782, Spain}
\newcommand{\ihep}{Institute of High Energy Physics, Chinese Academy of Sciences, Beijing, China}
\newcommand{\Iitk}{Indian Institute of Technology Kanpur, Uttar Pradesh 208016, India}
\newcommand{\Illinoisinstitute}{Illinois Institute of Technology, Chicago, IL 60616, USA}
\newcommand{\Imperial}{Imperial College of Science, Technology and Medicine, London SW7 2BZ, United Kingdom}
\newcommand{\IndGuwahati}{Indian Institute of Technology Guwahati, Guwahati, 781 039, India}
\newcommand{\IndHyderabad}{Indian Institute of Technology Hyderabad, Hyderabad, 502285, India}
\newcommand{\Indiana}{Indiana University, Bloomington, IN 47405, USA}
\newcommand{\INFNBologna}{Istituto Nazionale di Fisica Nucleare Sezione di Bologna, 40127 Bologna BO, Italy}
\newcommand{\INFNCatania}{Istituto Nazionale di Fisica Nucleare Sezione di Catania, I-95123 Catania, Italy}
\newcommand{\INFNFerrara}{Istituto Nazionale di Fisica Nucleare Sezione di Ferrara, I-44122 Ferrara, Italy}
\newcommand{\INFNFrascati}{Istituto Nazionale di Fisica Nucleare Laboratori Nazionali di Frascati, Frascati, Roma, Italy}
\newcommand{\INFNGenova}{Istituto Nazionale di Fisica Nucleare Sezione di Genova, 16146 Genova GE, Italy}
\newcommand{\INFNLecce}{Istituto Nazionale di Fisica Nucleare Sezione di Lecce, 73100 - Lecce, Italy}
\newcommand{\INFNMilanBicocca}{Istituto Nazionale di Fisica Nucleare Sezione di Milano Bicocca, 3 - I-20126 Milano, Italy}
\newcommand{\INFNMilano}{Istituto Nazionale di Fisica Nucleare Sezione di Milano, 20133 Milano, Italy}
\newcommand{\INFNNapoli}{Istituto Nazionale di Fisica Nucleare Sezione di Napoli, I-80126 Napoli, Italy}
\newcommand{\INFNPadova}{Istituto Nazionale di Fisica Nucleare Sezione di Padova, 35131 Padova, Italy}
\newcommand{\INFNPavia}{Istituto Nazionale di Fisica Nucleare Sezione di Pavia,  I-27100 Pavia, Italy}
\newcommand{\INFNPisa}{Istituto Nazionale di Fisica Nucleare Laboratori Nazionali di Pisa, Pisa PI, Italy}
\newcommand{\INFNRoma}{Istituto Nazionale di Fisica Nucleare Sezione di Roma, 00185 Roma RM, Italy}
\newcommand{\INFNRomavergata}{Istituto Nazionale di Fisica Nucleare Roma Tor Vergata , 00133 Roma RM, Italy}
\newcommand{\INFNSud}{Istituto Nazionale di Fisica Nucleare Laboratori Nazionali del Sud, 95123 Catania, Italy}
\newcommand{\Infntorino}{Istituto Nazionale di Fisica Nucleare, Sezione di Torino, Turin, Italy}
\newcommand{\Ingenieria}{Universidad Nacional de Ingenier\'ia, Lima 25, Per\'u}
\newcommand{\Insubria }{University of Insubria, Via Ravasi, 2, 21100 Varese VA, Italy}
\newcommand{\Iowa}{University of Iowa, Iowa City, IA 52242, USA}
\newcommand{\IowaState}{Iowa State University, Ames, Iowa 50011, USA}
\newcommand{\IPLyon}{Institut de Physique des 2 Infinis de Lyon, 69622 Villeurbanne, France}
\newcommand{\IPM}{Institute for Research in Fundamental Sciences, Tehran, Iran}
\newcommand{\IRLPPC}{Particle Physics and Cosmology International Research Laboratory	, Chicago IL,  60637 USA}
\newcommand{\ISTlisboa}{Instituto Superior T\'ecnico - IST, Universidade de Lisboa, 1049-001 Lisboa, Portugal}
\newcommand{\Ita}{Instituto Tecnol\'ogico de Aeron\'autica, Sao Jose dos Campos, Brazil}
\newcommand{\Iwate}{Iwate University, Morioka, Iwate 020-8551, Japan}
\newcommand{\Jacksonstate}{Jackson State University, Jackson, MS 39217, USA}
\newcommand{\Jawaharlal}{Jawaharlal Nehru University, New Delhi 110067, India}
\newcommand{\Jeonbuk}{Jeonbuk National University, Jeonrabuk-do 54896, South Korea}
\newcommand{\Jyvaskyla}{Jyv\"askyl\"a University, FI-40014 Jyv\"askyl\"a, Finland}
\newcommand{\Kansasstate}{Kansas State University, Manhattan, KS 66506, USA}
\newcommand{\Kavli}{Kavli Institute for the Physics and Mathematics of the Universe, Kashiwa, Chiba 277-8583, Japan}
\newcommand{\KEK}{High Energy Accelerator Research Organization (KEK), Ibaraki, 305-0801, Japan}
\newcommand{\KISTI}{Korea Institute of Science and Technology Information, Daejeon, 34141, South Korea}
\newcommand{\Kyiv}{Taras Shevchenko National University of Kyiv, 01601 Kyiv, Ukraine}
\newcommand{\Lancaster}{Lancaster University, Lancaster LA1 4YB, United Kingdom}
\newcommand{\LawrenceBerkeley}{Lawrence Berkeley National Laboratory, Berkeley, CA 94720, USA}
\newcommand{\LIP}{Laborat\'orio de Instrumenta{\c c}\~ao e F\'isica Experimental de Part\'iculas, 1649-003 Lisboa and 3004-516 Coimbra, Portugal}
\newcommand{\Liverpool}{University of Liverpool, L69 7ZE, Liverpool, United Kingdom}
\newcommand{\LosAlmos}{Los Alamos National Laboratory, Los Alamos, NM 87545, USA}
\newcommand{\Louisanastate}{Louisiana State University, Baton Rouge, LA 70803, USA}
\newcommand{\LpBordeaux}{Laboratoire de Physique des Deux Infinis Bordeaux - IN2P3, F-33175 Gradignan, Bordeaux, France, }
\newcommand{\Lucknow}{University of Lucknow, Uttar Pradesh 226007, India}
\newcommand{\Mainz}{Johannes Gutenberg-Universit\"at Mainz, 55122 Mainz, Germany}
\newcommand{\Manchester}{University of Manchester, Manchester M13 9PL, United Kingdom}
\newcommand{\Massinsttech}{Massachusetts Institute of Technology, Cambridge, MA 02139, USA}
\newcommand{\Medellin}{University of Medell\'in, Medell\'in, 050026 Colombia }
\newcommand{\Michigan}{University of Michigan, Ann Arbor, MI 48109, USA}
\newcommand{\Michiganstate}{Michigan State University, East Lansing, MI 48824, USA}
\newcommand{\MilanoBicocca}{Universit\`a di Milano Bicocca , 20126 Milano, Italy}
\newcommand{\MilanoUniv}{Universit\`a degli Studi di Milano, I-20133 Milano, Italy}
\newcommand{\Minnduluth}{University of Minnesota Duluth, Duluth, MN 55812, USA}
\newcommand{\Minntwin}{University of Minnesota Twin Cities, Minneapolis, MN 55455, USA}
\newcommand{\Mississippi}{University of Mississippi, University, MS 38677 USA}
\newcommand{\napoli}{Universit\`a degli Studi di Napoli Federico II , 80138 Napoli NA, Italy}
\newcommand{\Nikhef}{Nikhef National Institute of Subatomic Physics, 1098 XG Amsterdam, Netherlands}
\newcommand{\Niser}{National Institute of Science Education and Research (NISER), Odisha 752050, India}
\newcommand{\Northdakota}{University of North Dakota, Grand Forks, ND 58202-8357, USA}
\newcommand{\Northernillinois}{Northern Illinois University, DeKalb, IL 60115, USA}
\newcommand{\Northwestern}{Northwestern University, Evanston, Il 60208, USA}
\newcommand{\NotreDame}{University of Notre Dame, Notre Dame, IN 46556, USA}
\newcommand{\NoviSad}{University of Novi Sad, 21102 Novi Sad, Serbia}
\newcommand{\Ohiostate}{Ohio State University, Columbus, OH 43210, USA}
\newcommand{\OregonState}{Oregon State University, Corvallis, OR 97331, USA}
\newcommand{\Oxford}{University of Oxford, Oxford, OX1 3RH, United Kingdom}
\newcommand{\PacificNorthwest}{Pacific Northwest National Laboratory, Richland, WA 99352, USA}
\newcommand{\Padova}{Universt\`a degli Studi di Padova, I-35131 Padova, Italy}
\newcommand{\Panjab}{Panjab University, Chandigarh, 160014, India}
\newcommand{\Parissaclay}{Universit\'e Paris-Saclay, CNRS/IN2P3, IJCLab, 91405 Orsay, France}
\newcommand{\Parisuniversite}{Universit\'e Paris Cit\'e, CNRS, Astroparticule et Cosmologie, Paris, France}
\newcommand{\Parma}{University of Parma,  43121 Parma PR, Italy}
\newcommand{\Pavia}{Universit\`a degli Studi di Pavia, 27100 Pavia PV, Italy}
\newcommand{\Penn}{University of Pennsylvania, Philadelphia, PA 19104, USA}
\newcommand{\PennState}{Pennsylvania State University, University Park, PA 16802, USA}
\newcommand{\PhysicalResearchLaboratory}{Physical Research Laboratory, Ahmedabad 380 009, India}
\newcommand{\Pisa}{Universit\`a di Pisa, I-56127 Pisa, Italy}
\newcommand{\Pitt}{University of Pittsburgh, Pittsburgh, PA 15260, USA}
\newcommand{\Pontificia}{Pontificia Universidad Cat\'olica del Per\'u, Lima, Per\'u}
\newcommand{\PuertoRico}{University of Puerto Rico, Mayaguez 00681, Puerto Rico, USA}
\newcommand{\Punjab}{Punjab Agricultural University, Ludhiana 141004, India}
\newcommand{\QMUL}{Queen Mary University of London, London E1 4NS, United Kingdom
}
\newcommand{\Radboud}{Radboud University, NL-6525 AJ Nijmegen, Netherlands}
\newcommand{\Rice}{Rice University, Houston, TX 77005}
\newcommand{\Rochester}{University of Rochester, Rochester, NY 14627, USA}
\newcommand{\Royalholloway}{Royal Holloway College London, London, TW20 0EX, United Kingdom}
\newcommand{\Rutgers}{Rutgers University, Piscataway, NJ, 08854, USA}
\newcommand{\Rutherford}{STFC Rutherford Appleton Laboratory, Didcot OX11 0QX, United Kingdom}
\newcommand{\Salento}{Universit\`a del Salento, 73100 Lecce, Italy}
\newcommand{\santamarta}{Universidad del Magdalena, Santa Marta - Colombia}
\newcommand{\Sapienza}{Sapienza University of Rome, 00185 Roma RM, Italy}
\newcommand{\SergioArboleda}{Universidad Sergio Arboleda, 11022 Bogot\'a, Colombia}
\newcommand{\Sheffield}{University of Sheffield, Sheffield S3 7RH, United Kingdom}
\newcommand{\SLAC}{SLAC National Accelerator Laboratory, Menlo Park, CA 94025, USA}
\newcommand{\Southcarolina}{University of South Carolina, Columbia, SC 29208, USA}
\newcommand{\SouthDakotaSchool}{South Dakota School of Mines and Technology, Rapid City, SD 57701, USA}
\newcommand{\SouthDakotaState}{South Dakota State University, Brookings, SD 57007, USA}
\newcommand{\StonyBrook}{Stony Brook University, SUNY, Stony Brook, NY 11794, USA}
\newcommand{\SURF}{Sanford Underground Research Facility, Lead, SD, 57754, USA}
\newcommand{\Sussex}{University of Sussex, Brighton, BN1 9RH, United Kingdom}
\newcommand{\Syracuse}{Syracuse University, Syracuse, NY 13244, USA}
\newcommand{\Tecnologica }{Universidade Tecnol\'ogica Federal do Paran\'a, Curitiba, Brazil}
\newcommand{\TelAviv}{Tel Aviv University, Tel Aviv-Yafo, Israel}
\newcommand{\TexasAMcollege}{Texas A\&M University, College Station, Texas 77840}
\newcommand{\TexasAMcorpuscristi}{Texas A\&M University - Corpus Christi, Corpus Christi, TX 78412, USA}
\newcommand{\TexasArlington}{University of Texas at Arlington, Arlington, TX 76019, USA}
\newcommand{\Texasaustin}{University of Texas at Austin, Austin, TX 78712, USA}
\newcommand{\Toronto}{University of Toronto, Toronto, Ontario M5S 1A1, Canada}
\newcommand{\Tufts}{Tufts University, Medford, MA 02155, USA}
\newcommand{\Unifesp}{Universidade Federal de S\~ao Paulo, 09913-030, S\~ao Paulo, Brazil}
\newcommand{\UNIST}{Ulsan National Institute of Science and Technology, Ulsan 689-798, South Korea}
\newcommand{\UniversityCollegeLondon}{University College London, London, WC1E 6BT, United Kingdom}
\newcommand{\univkansas}{University of Kansas, Lawrence, KS 66045}
\newcommand{\UNMSM}{Universidad Nacional Mayor de San Marcos, Lima, Peru}
\newcommand{\ValleyCity}{Valley City State University, Valley City, ND 58072, USA}
\newcommand{\Vigo}{University of Vigo, E- 36310 Vigo Spain}
\newcommand{\VirginiaTech}{Virginia Tech, Blacksburg, VA 24060, USA}
\newcommand{\Warsaw}{University of Warsaw, 02-093 Warsaw, Poland}
\newcommand{\Warwick}{University of Warwick, Coventry CV4 7AL, United Kingdom}
\newcommand{\Wellesley}{Wellesley College, Wellesley, MA 02481, USA}
\newcommand{\Wichita}{Wichita State University, Wichita, KS 67260, USA}
\newcommand{\WilliamMary}{William and Mary, Williamsburg, VA 23187, USA}
\newcommand{\Wisconsin}{University of Wisconsin Madison, Madison, WI 53706, USA}
\newcommand{\Yale}{Yale University, New Haven, CT 06520, USA}
\newcommand{\Yerevan}{Yerevan Institute for Theoretical Physics and Modeling, Yerevan 0036, Armenia}
\newcommand{\York}{York University, Toronto M3J 1P3, Canada}
\affiliation{\Albanysuny}
\affiliation{\Almaty}
\affiliation{\Amsterdam}
\affiliation{\Antalya}
\affiliation{\Antananarivo}
\affiliation{\Antioquia}
\affiliation{\AntonioNarino}
\affiliation{\Argonne}
\affiliation{\Arizona}
\affiliation{\Asuncion}
\affiliation{\Athens}
\affiliation{\Atlantico}
\affiliation{\Augustana}
\affiliation{\Bern}
\affiliation{\Beykent}
\affiliation{\Birmingham}
\affiliation{\BolognaUniversity}
\affiliation{\Boston}
\affiliation{\Bristol}
\affiliation{\Brookhaven}
\affiliation{\Bucharest}
\affiliation{\CalBerkeley}
\affiliation{\CalDavis}
\affiliation{\CalIrvine}
\affiliation{\CalLosangeles}
\affiliation{\CalRiverside}
\affiliation{\CalSantabarbara}
\affiliation{\Caltech}
\affiliation{\Cambridge}
\affiliation{\Campinas}
\affiliation{\CataniaUniversitadi}
\affiliation{\Catolica}
\affiliation{\CBPF}
\affiliation{\CEASaclay}
\affiliation{\CERN}
\affiliation{\Charles}
\affiliation{\Chicago}
\affiliation{\ChungAng}
\affiliation{\CIEMAT}
\affiliation{\Cincinnati}
\affiliation{\Cinvestav}
\affiliation{\Colima}
\affiliation{\ColoradoBoulder}
\affiliation{\ColoradoState}
\affiliation{\Columbia}
\affiliation{\conida}
\affiliation{\Cti}
\affiliation{\CUSB}
\affiliation{\CzechAcademyofSciences}
\affiliation{\CzechTechnical}
\affiliation{\DannecyleVieux}
\affiliation{\Daresbury}
\affiliation{\Dordt}
\affiliation{\Drexel}
\affiliation{\Duke}
\affiliation{\Durham}
\affiliation{\Edinburgh}
\affiliation{\EIA}
\affiliation{\Eotvos}
\affiliation{\erciyes}
\affiliation{\FCULport}
\affiliation{\FederaldeAlfenas}
\affiliation{\FederaldeGoias}
\affiliation{\FederaldoABC}
\affiliation{\FederaldoRio}
\affiliation{\Fermi}
\affiliation{\Ferrarauniv}
\affiliation{\Florida}
\affiliation{\Floridastate}
\affiliation{\Fluminense}
\affiliation{\Genova}
\affiliation{\Georgian}
\affiliation{\Granada}
\affiliation{\GranSasso}
\affiliation{\GranSassoLab}
\affiliation{\Grenoble}
\affiliation{\Guanajuato}
\affiliation{\Harish}
\affiliation{\Hawaii}
\affiliation{\hkust}
\affiliation{\Houston}
\affiliation{\Hyderabad}
\affiliation{\Idaho}
\affiliation{\IFIC}
\affiliation{\IGFAE}
\affiliation{\ihep}
\affiliation{\Iitk}
\affiliation{\Illinoisinstitute}
\affiliation{\Imperial}
\affiliation{\IndGuwahati}
\affiliation{\IndHyderabad}
\affiliation{\Indiana}
\affiliation{\INFNBologna}
\affiliation{\INFNCatania}
\affiliation{\INFNFerrara}
\affiliation{\INFNFrascati}
\affiliation{\INFNGenova}
\affiliation{\INFNLecce}
\affiliation{\INFNMilanBicocca}
\affiliation{\INFNMilano}
\affiliation{\INFNNapoli}
\affiliation{\INFNPadova}
\affiliation{\INFNPavia}
\affiliation{\INFNPisa}
\affiliation{\INFNRoma}
\affiliation{\INFNRomavergata}
\affiliation{\INFNSud}
\affiliation{\Infntorino}
\affiliation{\Ingenieria}
\affiliation{\Insubria }
\affiliation{\Iowa}
\affiliation{\IowaState}
\affiliation{\IPLyon}
\affiliation{\IPM}
\affiliation{\IRLPPC}
\affiliation{\ISTlisboa}
\affiliation{\Ita}
\affiliation{\Iwate}
\affiliation{\Jacksonstate}
\affiliation{\Jawaharlal}
\affiliation{\Jeonbuk}
\affiliation{\Jyvaskyla}
\affiliation{\Kansasstate}
\affiliation{\Kavli}
\affiliation{\KEK}
\affiliation{\KISTI}
\affiliation{\Kyiv}
\affiliation{\Lancaster}
\affiliation{\LawrenceBerkeley}
\affiliation{\LIP}
\affiliation{\Liverpool}
\affiliation{\LosAlmos}
\affiliation{\Louisanastate}
\affiliation{\LpBordeaux}
\affiliation{\Lucknow}
\affiliation{\Mainz}
\affiliation{\Manchester}
\affiliation{\Massinsttech}
\affiliation{\Medellin}
\affiliation{\Michigan}
\affiliation{\Michiganstate}
\affiliation{\MilanoBicocca}
\affiliation{\MilanoUniv}
\affiliation{\Minnduluth}
\affiliation{\Minntwin}
\affiliation{\Mississippi}
\affiliation{\napoli}
\affiliation{\Nikhef}
\affiliation{\Niser}
\affiliation{\Northdakota}
\affiliation{\Northernillinois}
\affiliation{\Northwestern}
\affiliation{\NotreDame}
\affiliation{\NoviSad}
\affiliation{\Ohiostate}
\affiliation{\OregonState}
\affiliation{\Oxford}
\affiliation{\PacificNorthwest}
\affiliation{\Padova}
\affiliation{\Panjab}
\affiliation{\Parissaclay}
\affiliation{\Parisuniversite}
\affiliation{\Parma}
\affiliation{\Pavia}
\affiliation{\Penn}
\affiliation{\PennState}
\affiliation{\PhysicalResearchLaboratory}
\affiliation{\Pisa}
\affiliation{\Pitt}
\affiliation{\Pontificia}
\affiliation{\PuertoRico}
\affiliation{\Punjab}
\affiliation{\QMUL}
\affiliation{\Radboud}
\affiliation{\Rice}
\affiliation{\Rochester}
\affiliation{\Royalholloway}
\affiliation{\Rutgers}
\affiliation{\Rutherford}
\affiliation{\Salento}
\affiliation{\santamarta}
\affiliation{\Sapienza}
\affiliation{\SergioArboleda}
\affiliation{\Sheffield}
\affiliation{\SLAC}
\affiliation{\Southcarolina}
\affiliation{\SouthDakotaSchool}
\affiliation{\SouthDakotaState}
\affiliation{\StonyBrook}
\affiliation{\SURF}
\affiliation{\Sussex}
\affiliation{\Syracuse}
\affiliation{\Tecnologica }
\affiliation{\TelAviv}
\affiliation{\TexasAMcollege}
\affiliation{\TexasAMcorpuscristi}
\affiliation{\TexasArlington}
\affiliation{\Texasaustin}
\affiliation{\Toronto}
\affiliation{\Tufts}
\affiliation{\Unifesp}
\affiliation{\UNIST}
\affiliation{\UniversityCollegeLondon}
\affiliation{\univkansas}
\affiliation{\UNMSM}
\affiliation{\ValleyCity}
\affiliation{\Vigo}
\affiliation{\VirginiaTech}
\affiliation{\Warsaw}
\affiliation{\Warwick}
\affiliation{\Wellesley}
\affiliation{\Wichita}
\affiliation{\WilliamMary}
\affiliation{\Wisconsin}
\affiliation{\Yale}
\affiliation{\Yerevan}
\affiliation{\York}
\author{S.~Abbaslu} \affiliation{\IPM}
\author{F.~Abd Alrahman} \affiliation{\Houston}
\author{A.~Abed Abud} \affiliation{\CERN}
\author{R.~Acciarri} \affiliation{\CERN}
\author{L.~P.~Accorsi} \affiliation{\Tecnologica }
\author{M.~A.~Acero} \affiliation{\Atlantico}
\author{M.~R.~Adames} \affiliation{\Tecnologica }
\author{G.~Adamov} \affiliation{\Georgian}
\author{M.~Adamowski} \affiliation{\Fermi}
\author{C.~Adriano} \affiliation{\Campinas}
% ...
\collaboration{The DUNE Collaboration}
\noaffiliation

\date{\today}%

\begin{abstract}
\clearpage

\noindent The Deep Underground Neutrino Experiment (DUNE) is a next-generation neutrino experiment with a rich physics program that includes searches for the hypothetical phenomenon of proton decay. Utilizing liquid-argon time-projection chamber technology, DUNE is expected to achieve world-leading sensitivity in the proton decay channels that involve charged kaons in their final states. The first DUNE demonstrator, ProtoDUNE Single-Phase, was a 0.77~kt detector that operated from 2018 to 2020 at the CERN Neutrino Platform, exposed to a mixed hadron and electron test-beam with momenta ranging from 0.3 to 7~GeV/c. We present a selection of low-energy kaons among the secondary particles produced in hadronic reactions, using data from the 6 and 7~GeV/c beam runs. The selection efficiency is 1\% and the sample purity 92\%. The initial energies of the selected kaon candidates encompass the expected energy range of kaons originating from proton decay events in DUNE (below $\sim$200~MeV). In addition, we demonstrate the capability of this detector technology to discriminate between kaons and other particles such as protons and muons, and provide a comprehensive description of their energy loss in liquid argon, which shows good agreement with the simulation. These results pave the way for future proton decay searches at DUNE.
\end{abstract}

\maketitle

\section{INTRODUCTION}
\noindent The Standard Model (SM) of particle physics is the most successful and precise framework for describing particle interactions and has been confirmed by numerous experimental observations. However, it remains an incomplete theory of nature. In particular, it fails to fully explain key fundamental phenomena, including the neutrino masses, the matter-antimatter asymmetry of the universe, the quantization of electric charge, and the existence of dark matter~\cite{bib:sm}. Grand Unification Theories (GUTs) extend the SM to predict the unification of electromagnetic, weak and strong interactions at very high energies ($\sim10^{16}$GeV)~\cite{bib:guts,bib:guts_2}, presenting a rich phenomenology that can lead to solutions for the issues above~\cite{bib:gut_pheno_1,bib:gut_pheno_2}. Most GUTs allow leptons and quarks to transform into each other, inducing baryon-number violation processes with very long lifetimes such as proton decay~\cite{bib:protdecay_1,bib:protdecay_2,bib:protdecay_3}, an observation of which would constitute an experimental proof of GUTs. 

Proton lifetime and decay channels depend on the GUT model. The most favoured decay channels are $p \rightarrow l^{+}\pi^{0}$ ($l=e,\mu$) and $p \rightarrow K^{+}\bar{\nu}$~\cite{bib:protdecay_1}. Due to its large fiducial mass and long data-taking period, the Super-Kamiokande (SK) water Cherenkov detector has been able to set the strongest upper limits for several channels. In the case of channels with a pion in the final state, SK set a limit at 90\% CL of $1.6\times10^{34}$ years for $p \rightarrow \mu^{+}\pi^{0}$, and $2.4\times10^{34}$ years for $p \rightarrow e^{+}\pi^{0}$ \cite{bib:sk_pdecay1}. SK has also set a limit on $p \rightarrow K^{+}\bar{\nu}$ of $5.9\times10^{33}$ years~\cite{bib:sk_pdecay2}, which is not as strong as in the previous channels. This is due to the fact that, in this two-body decay of a stationary proton, the $K^{+}$ momentum (340 MeV/c) is below the Cherenkov threshold in SK, and therefore it can only be detected via the decay products of the kaon. On the contrary, DUNE~\cite{bib:dune_tdr1} has the capability to reconstruct the full decay chain---kaon and its decay products---for this process by exploiting the Liquid-Argon Time Projection Chamber (LArTPC) technology.

DUNE is a future long-baseline neutrino experiment aiming to provide the most comprehensive study of the neutrino oscillation phenomenon. It consists of a powerful neutrino beam, a Near Detector (ND) complex to characterize the initial, unoscillated neutrino flux~\cite{bib:nd_cdr}, and a Far Detector (FD) complex to measure the effect of neutrino oscillations~\cite{bib:dune_tdr1}. The FD will be located 1.5~km underground and 1300 km away from the ND at the Sanford Underground Research Facility, which will provide a powerful shielding against cosmic rays. This, combined with its large fiducial mass ($\sim$40~kt), makes DUNE FD sensitive to different rare events such as baryon-number-violating processes, particularly to proton decay channels with positively-charged kaons in the final state \cite{bib:dune_tdr2,bib:dune_bsm}. 

The most likely decay channel for charged kaons is $K^{+} \rightarrow \mu^{+}\nu_{\mu}$ (64\% branching ratio). 
The proton decay signature would consist of a low momentum kaon ($p_{K} \lesssim 340$ MeV/c due to intranuclear effects) originating in the active volume of the TPC followed by a low momentum muon ($p_{\mu} \sim 237$ MeV/c, due to the two-body decay kinematics). The dominant background for this process is generated by atmospheric neutrino charged-current quasi-elastic scattering, $\nu_{\mu}n \rightarrow p\mu^{-}$, as it present a similar topology to that of $K^{+}\mu^{+}$ (a vertex originating in the active volume of the detector with a muon track and a short, highly ionizing track corresponding to the proton). If the muon has a momentum similar to the 237 MeV/c expected from the $K^{+}$ decay, the identification of the proton decay process depends on the ability to differentiate between kaons and protons. DUNE sensitivity studies have shown that a 30\% signal efficiency is feasible. This, combined with an expected background acceptance of one event per Mt$\cdot$year, can yield a 90\% CL lower limit on the proton lifetime in the $p \rightarrow K^{+} \bar{\nu}$ channel of $1.3 \times 10^{34}$ years, assuming no signal is observed over ten years of running with a total of 40~kt of fiducial mass \cite{bib:dune_tdr2}.

As a first step towards proton decay searches in DUNE, this work uses ProtoDUNE Single-Phase (SP) data to demonstrate the capabilities of the LArTPC technology to identify low-energy kaons ($p_{K}\lesssim340$ MeV/c) and to characterize their energy loss in liquid argon. The ProtoDUNE-SP detector is described in detail in Sec. \ref{sec:pdsp}, and Sec. \ref{sec:simreco} outlines the simulation and reconstruction of ProtoDUNE-SP data. Sec. \ref{sec:selection} presents the event selection developed to isolate low-energy kaons in the detector and Sec. \ref{sec:fit} describes the methodology to evaluate their energy loss in LAr. Sec. \ref{sec:syst} evaluates the systematic uncertainties considered in this analysis, Sec. \ref{sec:results} shows the obtained results and Sec. \ref{sec:conclusion} concludes.

\section{THE PROTODUNE-SP DETECTOR AT CERN}
\label{sec:pdsp}

\noindent ProtoDUNE-SP was the first DUNE FD demonstrator~\cite{bib:pdsp_tdr}. Located at CERN's Neutrino Platform~\cite{bib:cern_np_1,bib:cern_np_2}, it was a LArTPC experiment which prototyped full-sized DUNE's FD single-phase\footnote{Now Horizontal Drift (HD).} detector components using about a 5\% of the of the DUNE FD argon mass. It constituted the largest monolithic LArTPC ever built and operated, with a total LAr mass of 0.77~kt. Its installation was completed in Summer 2018, and it was commissioned and exposed to a test-beam from September to November 2018. The beam provided particles of different types and momenta so that the detector could be calibrated and its performance validated~\cite{bib:pdsp_firstresults,bib:pdsp_performance}.

\begin{figure}[htbp]
\centering
\includegraphics[width=0.45\textwidth]{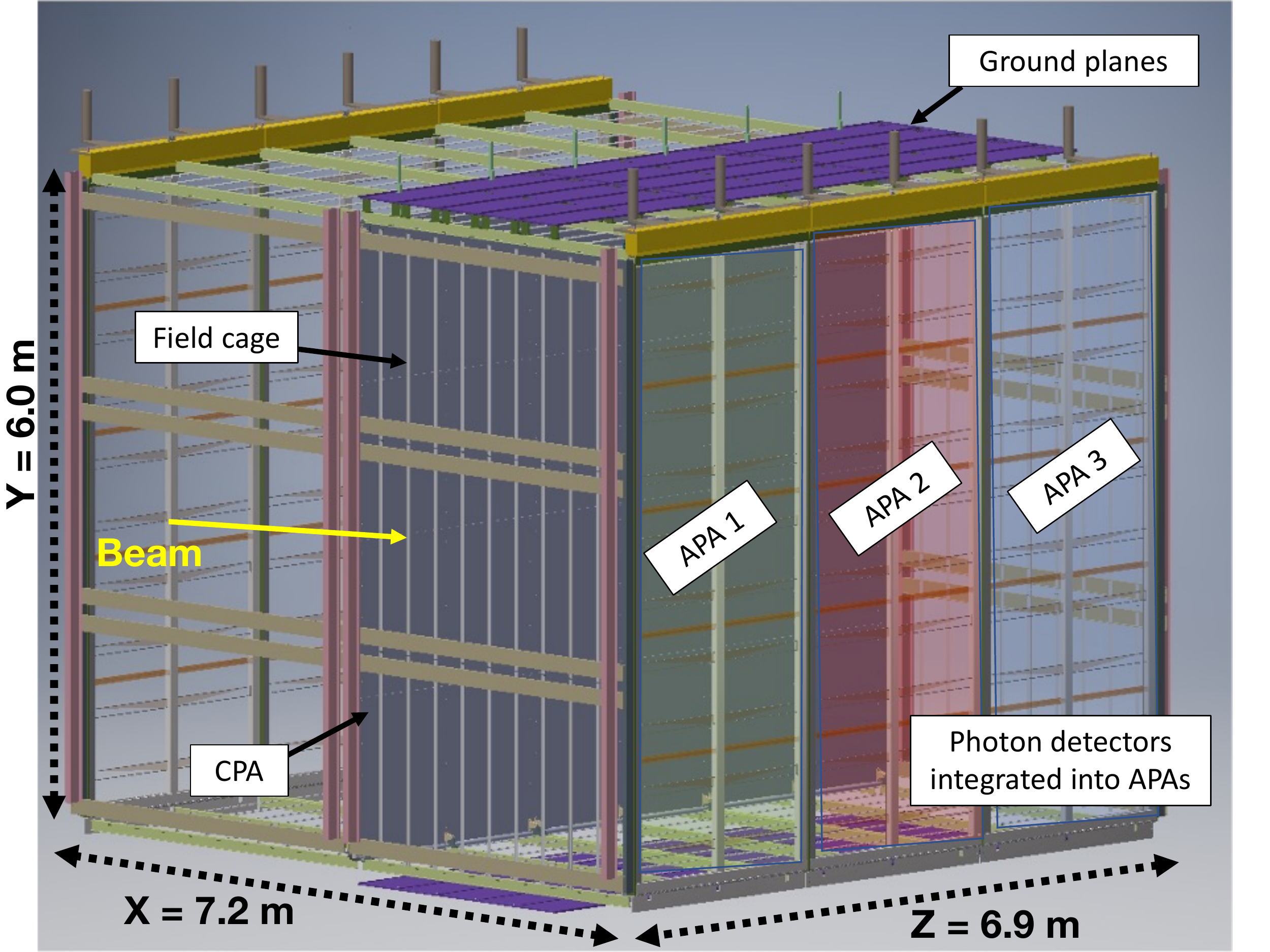}
\caption{Configuration of ProtoDUNE-SP LArTPC. CPA refers to Cathode Plane Assembly.
\label{fig:detector}}
\end{figure}

ProtoDUNE-SP was housed in a cryostat with outer dimensions of 11.4~m$\times$10.8$\times$11.4~m$^{3}$ (X$\times$Y$\times$Z) that kept the LAr at 87~K. A diagram of the detector can be seen in Fig.~\ref{fig:detector}. The TPC had active dimensions of 7.2~m$\times$6.0~m$\times$6.9~m, separated into two volumes by a central cathode plane located at $x=0$ and parallel to the Z axis. The Y axis was vertical. Each volume had a drift length of 3.6~m and was exposed to an electric field of 500~V/cm, generated by a -180~kV bias applied to the shared cathode plane. The electric field homogeneity was ensured by a field cage enclosing the active volume on the non-anode faces of the detector. The anode planes were formed of three Anode Plane Assemblies (APA) each, supported by a stainless-steel frame of 6.1~m high, 2.3~m wide and 76~mm thick. Four copper-beryllium wire planes were bonded directly over both sides of the APA frame, the first one of which served as ground plane (G), the next two were induction planes (U and V), and the last one was a collection plane (X). U wires were set at 35.7$^{\circ}$ with respect to the vertical axis, V wires at -35.7$^{\circ}$, and X and G wires parallel to it. The selected wire angles made induction wires cross only once a given collection wire, reducing reconstruction ambiguities. Voltages of the wire planes were staggered ($V_{G} = -665$~V, $V_{U} = -370$~V, $V_{V} = 0$~V and $V_{X} = 820$~V) such that electrons could flow through the induction wires towards the collection plane. Photon detection modules were embedded in the stainless-steel frame, behind the highly transparent wire planes. Electron diverters were installed in the non-sensitive gap between APAs to reduce the amount of charge reaching this region. These were formed by two electrode strips mounted on an insulating board, which after being biased by an external voltage, modified the local drift field in such a way that electrons were redirected towards the active regions of the APAs and away from the gaps \cite{bib:pdsp_tdr}. During the operation of the detector, the electron diverters developed short-circuits that produced high currents in the active volume of the detector, and were therefore grounded. This caused a loss of collected charge that ultimately led to reconstruction errors, in which a single track crossing two APAs was reconstructed as two separated tracks, with the second segment incorrectly associated as a daughter of the first.~\cite{bib:pdsp_firstresults,bib:pdsp_performance}.  

In order to minimize the energy loss of particles from the test-beam crossing the cryostat, a beam plug was installed at the front face of the detector at the beam entry point. It consisted of a cylindrical pressure vessel filled with nitrogen gas that penetrated inside the TPC (up to five centimeters inside the field cage), so that the interaction of beam particles and inactive material was minimized. The beam entry point was approximately located at mid-height and about 30 cm away from the cathode. It was roughly aligned with the positive-z direction, deviating by approximately 15$^{\circ}$ towards the negative-x and negative-y directions, in such a way that beam particles crossed only one drift volume~\cite{bib:pdsp_tdr}. 

The test-beam was provided by the CERN H4 beamline. The proton beam extracted from the Super Proton Synchrotron (SPS) is shot onto a beryllium target generating a mixed hadron beam of about 80~GeV/c of momentum. This secondary beam is sent once more towards another target (made of copper or tungsten), providing a mixed tertiary beam of protons, positrons, and positively charged kaons, pions and muons, the momenta of which was set at 0.3, 0.5, 1, 2, 3, 6 and 7~GeV/c by downstream magnets. The tungsten target is used to increase the hadron content of the beam at low momentum (below 4 GeV/c), however it was unintentionally not used for the 2~GeV/c runs, substantially decreasing the hadron abundance of the 2~GeV/c runs. The tertiary beam, instrumented with a sequence of monitors for particle identification~\cite{bib:pdsp_beamline_1,bib:pdsp_beamline_2}, provided test-beam particles with a frequency of 25~Hz, which allowed particles to be characterized one-at-a-time as the TPC readout time window was 5~ms~\cite{bib:pdsp_tdr}.

As a surface detector, ProtoDUNE-SP was exposed to a high flux of cosmic rays which continuously crossed the detector thereby ionizing argon atoms. The ionization electrons drift towards the anode at a velocity of $\sim$1.6~mm/\textmu s, whereas the more massive Ar$^{+}$ ions drift towards to the cathode at a much smaller velocity ($\sim$5~mm/s) \cite{bib:reco_1}. Slow drifting ions lead to build up of positive charge in the active volume, which modified the electric field lines inside the TPC. This so-called Space Charge Effect (SCE)~\cite{bib:sce} had a significant impact on ProtoDUNE-SP spatial and calorimetric reconstruction due to the large dimensions of the detector \cite{bib:pdsp_firstresults}. SCE calibration and simulation was performed by measuring cosmic-ray tracks distortion near the boundaries of the active volume, where the SCE was maximal. However, significant differences were still observed between data and simulation in the boundaries of the active volume, including the front face of the detector where the test-beam particles were injected. Thus, a fiducial volume cut excluding the first 33 cm of the TPC on the Z axis has been adopted by ProtoDUNE-SP analyses to avoid systematic differences between data and simulation~\cite{bib:pdsp_kxs}. The high flux of cosmic rays also significantly limited the use of the photon detection system in high level physics analysis.

\section{SIMULATION, RECONSTRUCTION AND CALIBRATION}
\label{sec:simreco}
\noindent The simulation of particle transport and particle interactions in ProtoDUNE-SP is performed using GEANT4~\cite{bib:g4_1,bib:g4_2,bib:g4_3}, and encompasses the entire beamline facility. The primary particles considered are those originating from the beam itself, from the surrounding beam halo, and from cosmic rays. Beam and beam halo particles are generated using the G4Beamline event generator~\cite{bib:g4beamline}, FLUKA~\cite{bib:fluka_1,bib:fluka_2} and MAD-X~\cite{bib:madx}, while cosmic rays are simulated with CORSIKA~\cite{bib:corsika}. The propagation of these particles and the signal simulation was done within LArSoft~\cite{bib:larsoft} and it is described in detail in Ref.~\cite{bib:pdsp_firstresults}.

After an event is read out, the waveforms recorded by the induction and collection wires of the TPC are deconvolved and noise-filtered, and then analyzed to identify hits. A hit is an ionization charge deposited in space and time by a charged particle and detected by the three sensing planes of wires. These appear as peaks in the waveforms and are fitted to Gaussian functions to extract their characteristics. Using those hits as input, Pandora~\cite{bib:pandora_1,bib:pandora_2}, a multi-algorithm reconstruction package, is used to reconstruct the different particle interactions and the relationships between particles (namely, particle hierarchies). The hits are first processed separately in different drift volumes assuming the cosmic-ray hypothesis, and the reconstructed tracks are stitched together across volumes. Tracks that can be clearly ascribed to cosmic rays are tagged as such and put aside. The remaining hits are separated into groups, with the aim of containing all hits from the same interaction or cosmic ray in the same group. Each group of hits is reconstructed again with both the cosmic and the test-beam hypothesis, and the best outcome (cosmic or test-beam interaction) is then chosen via a boosted decision tree. The result of this process is a hierarchy of particles, in which daughters and parents are associated, and each particle is formed by a set of hits. A detailed description of the reconstruction is presented in Refs.~\cite{bib:pandora_2,bib:pdsp_firstresults,bib:pandora_microboone}. 

 The deposited energy per unit length ($\mathrm{d}E/\mathrm{d}x$) is the main handle used to identify particles in a LArTPC. The energy loss pattern along the trajectory is different for each particle type, especially near the Bragg peak region, where the energy deposition is maximum before the particle stops. In general, the larger the mass difference between particle types, the easier it is to differentiate them. Thus, it is essential to understand the relationship between the deposited energy and the response of the detector. A detailed description of the calorimetric calibration of the ProtoDUNE-SP TPC can be found in Ref.~\cite{bib:pdsp_firstresults}. The $\mathrm{d}E/\mathrm{d}x$ of each hit is computed by 

\begin{enumerate}
	\item Converting ADC counts to number of electrons. The pitch of the hit is computed based on the angle of the track with respect to the wire planes, providing the collected charge per unit length $\mathrm{d}Q/\mathrm{d}x$.
	\item Correcting the collected charge for the SCE and attenuation due to impurities, obtaining the deposited charge.
	\item Correcting the deposited charge for residual detector non-uniformities (such as non-working wires, electron diverters, etc.).
	\item Converting the deposited charge to deposited energy considering electron recombination effects \cite{bib:recombination_1,bib:recombination_2,bib:recombination_3,bib:recombination_4,bib:recombination_icarus}. In ProtoDUNE-SP, the recombination effect is described using the Modified Box Model~\cite{bib:recombination_argoneut}.
\end{enumerate}


\section{EVENT SELECTION}
\label{sec:selection} 

\noindent In ProtoDUNE-SP, most beam kaons of the 1~GeV/c beam runs decay before reaching the detector, while those of the 3~GeV/c beam runs or above typically undergo inelastic interactions when reaching the TPC. Hence, 2~GeV/c beam runs were considered optimal for producing stopping kaons. These could have been selected in relatively clean topologies, allowing for detailed studies of the energy profile and of the reconstruction efficiency as a function of the kaon's initial energy and/or other kinematic variables. Furthermore, tracks could have been artificially truncated to resemble those expected from proton decay events. Unfortunately, the tungsten target of the secondary beam was not used in the 2~GeV/c runs, significantly reducing the fraction of hadrons and making isolating primary stopping kaons more challenging.

As kaon production in hadronic interactions increases with the energy of the incoming hadron~\cite{bib:expected_k}, 6 and 7~GeV/c beam runs were therefore used in this study to isolate stopping---and potentially low-energy, $p_{K}\lesssim340$ MeV/c---kaons among the secondary particles generated in beam-hadron interactions in the active volume of the detector. However, even at these momenta, kaons are only marginally produced in hadronic interactions in comparison with other hadrons such as $\pi^{\pm}$ and protons, which dominate the backgrounds in this study and therefore require a very restrictive event selection. The dataset used in this analysis comprises 334882 and 269317 beam events at 6 and 7~GeV/c, respectively, of which 83.0\% are identified as $\pi^{+}/\mu^{+}/e^{+}$, 11.5\% as protons, and 5.5\% as kaons by the beam instrumentation. As $\pi^{+}/\mu^{+}/e^{+}$ cannot be distinguished by the beam instrumentation at these momenta, electron or muon beam events cannot be vetoed.  

The event selection is based on the kaon's main decay channel, which is $K^{+} \rightarrow \mu^{+} \nu_{\mu}$. The true signal is defined as kaons decaying via $K^{+} \rightarrow \mu^{+} \nu_{\mu}$, with a minimum kinetic energy threshold for the $K^{+}$ of 30 MeV---in order to account for detection thresholds. From the reconstruction point of view, as neutrinos are very unlikely to interact near the kaon decay point, the reconstructed signal is a track originating from a beam particle interaction, and with a single daughter compatible with a stopping muon of 237~MeV/c momentum (which is indeed the same signature that will be used in DUNE for proton decay searches). 

On average, each 6-7~GeV/c beam particle produces about 6 secondary particles, which in turn generate further particle production. Thus, multiple tracks can be compatible with the candidate topology in a single beam event. Consequently, the selection presented here does not work in a standard event-by-event basis, but in a candidate-by-candidate basis, meaning that more than one candidate in a beam event can pass the selection. We define the purity of the selection after the cut $i$ as

\begin{equation}
	Pur_{i} = \left.\frac{\textrm{Signal candidates}}{\textrm{Total candidates}}\right\vert_{i}\mathrm{,}
\end{equation}

\noindent where `Signal candidates' refers to the number of reconstructed candidates associated with true signal objects, and `Total candidates' refers to the total number of reconstructed candidates, both evaluated after cut $i$. Similarly, we define the efficiency of the selection after the cut $i$ as

\begin{equation}
	Eff_{i} = \left.\frac{\textrm{True signal candidates}}{\textrm{Total true signal candidates}}\right\vert_{i}\mathrm{,}
\end{equation}

\noindent where `True signal candidates' refers to the number of true signal events that are selected after cut $i$, and `Total true signal candidates' refers to the initial number of true signal objects. Purity and efficiency are both calculated using the simulation sample, and the selection has been optimized by maximizing $Pur\times Eff$ for each cut.

As previously mentioned, the energy loss profile of the tracks in the detector can be used for particle identification. Here we quantify the agreement between a reconstructed track and a particle hypothesis $Part$ by means of the parameter $\chi^{2}_{Part}/\mathrm{ndf}$, used in the selection and defined as:

\begin{equation}
	\chi^{2}_{Part}/\mathrm{ndf}=\frac{1}{N_{\mathrm{hits}}}\sum_{i}^{N_{\mathrm{hits}}}\frac{(\left.\frac{\mathrm{d}E}{\mathrm{d}x}\right|_{i}^{\mathrm{Reco}}-\left.\frac{\mathrm{d}E}{\mathrm{d}x}\right|_{i}^{\mathrm{Sim}})^{2}}{\sqrt{[\sigma(\left.\frac{\mathrm{d}E}{\mathrm{d}x}\right|_{i}^{\mathrm{Reco}})]^{2}+[\sigma(\left.\frac{\mathrm{d}E}{\mathrm{d}x}\right|_{i}^{\mathrm{Sim}})]^{2}}} \mathrm{,}
\end{equation}

\noindent where, $i$ runs over all the collection plane hits of the track within its last 26 cm\footnote{We use the last 26 cm of the tracks since the Bragg peak region of stopping particles is well contained there. For tracks longer than 26 cm, only the final 26 cm are considered, whereas shorter tracks are used in full.}, $\sigma(\frac{\mathrm{d}E}{\mathrm{d}x}|_{i})$ is the associated uncertainty of the $\frac{\mathrm{d}E}{\mathrm{d}x}$ for the $i$-th hit, `Reco' refers to the reconstructed track that is being evaluated, and `Sim' refers to the expectation based on the simulation for a given particle type. Events containing at least one potential kaon are selected following these steps:

\begin{itemize}
	\item The event must have a reconstructed beam particle inside the TPC.
	\item The event must have at least one particle compatible with the candidate kaon topology presented above. That is, a track (the kaon candidate) resulting from the interaction of a test-beam particle in the active volume of the detector, and with a single reconstructed daughter.
	\item The kaon candidate track must be fully contained within the fiducial volume, defined as the TPC region after the first 33 cm along the Z axis.
	\item The daughter of the kaon candidate track must
	\begin{itemize}
		\item be a track (not a shower, as these are generated by electrons and photons),
		\item have a reduced $\chi^{2}_{\mu}$ consistent with muon hypothesis, $\chi^2_{\mu}/\mathrm{ndf}<$ 6, and
		\item have a momentum derived from the track length and assuming the muon hypothesis between 221 and 245~MeV/c. The momentum is computed by applying the Continuous Slowing Down Approximation (CSDA)~\cite{bib:csda}.
	\end{itemize}
	\item The cosine of the angle between the kaon candidate track and its daughter must be smaller than 0.64. As stopping kaons decay at rest, the angular distribution of the muon with respect to the kaon is isotropic, while that of the daughters of other interacting hadrons is forward focused. Hence, rejecting the forward distribution is a powerful background rejection, similar to the approach followed in Ref.~\cite{bib:minerva_kinks}.
	\item The distance between the endpoint of the kaon candidate track and the starting point of its daughter must be less than 10~cm. This reduces the backgrounds originated from misreconstructions of inelastic interactions.
	\item The kaon candidate track must have a reduced $\chi^{2}_{K}$ consistent with kaon hypothesis, $\chi^2_{K}/\mathrm{ndf}<$ 50.
\end{itemize}

\begin{figure}[htbp]
\centering
\includegraphics[width=0.45\textwidth]{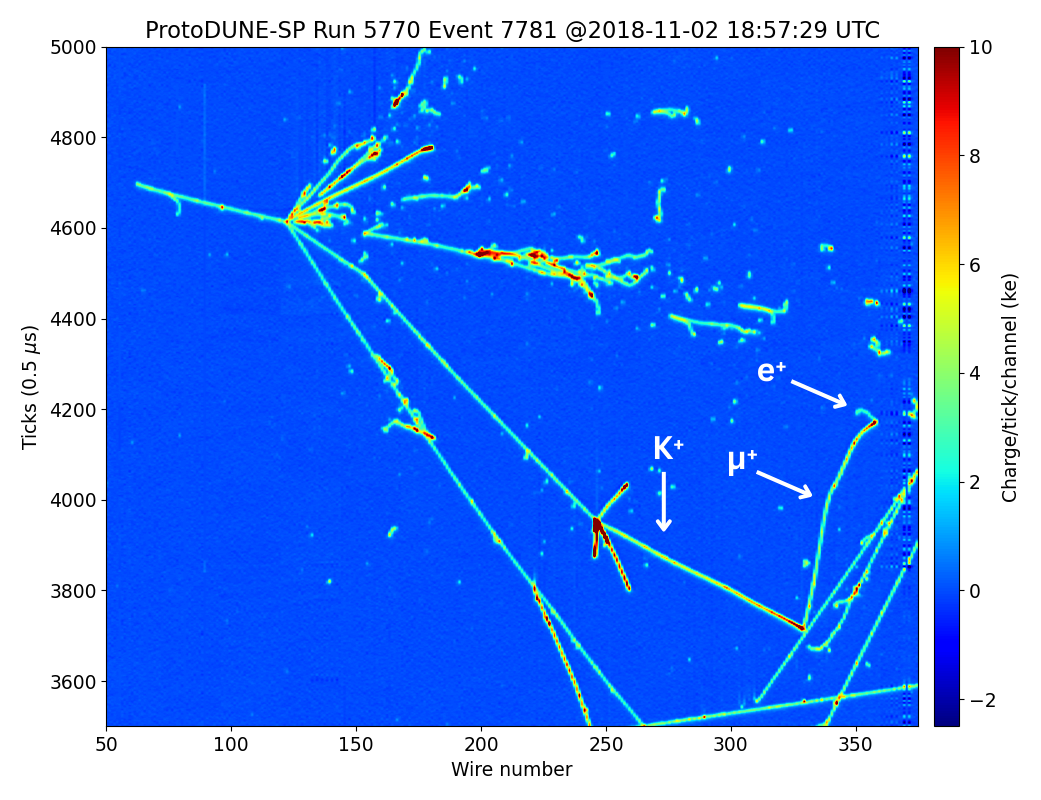}
\caption{Kaon candidate event display. A beam hadron track can be seen in the left side of the image. It interacts and generates secondary particles. One of these particles propagates in the argon until it interacts again, generating a kaon candidate. The white arrow points to the kaon track, which is followed by a muon track, ultimately followed by a Michel electron. 50 wires correspond approximately to 24~cm.
\label{fig:event_display}}
\end{figure}

The final selected data sample of 522 kaon candidates has a signal purity of 92\%. Table~\ref{tab:selectionresult} shows the evolution of the purity and efficiency for the different steps of the selection. For a true kaon to be selected it necessarily has to be associated to a reconstructed candidate, meaning that the kaon track and the muon track have to be reconstructed and associated as parent-daughter. This requirement, combined with the complicated topologies resulting from hadronic interactions as the one shown in Fig.~\ref{fig:event_display}, explains the significant drop in efficiency that is observed in the first step of Table \ref{tab:selectionresult}. 

\begin{figure}[htbp]
\centering
\includegraphics[width=0.45\textwidth]{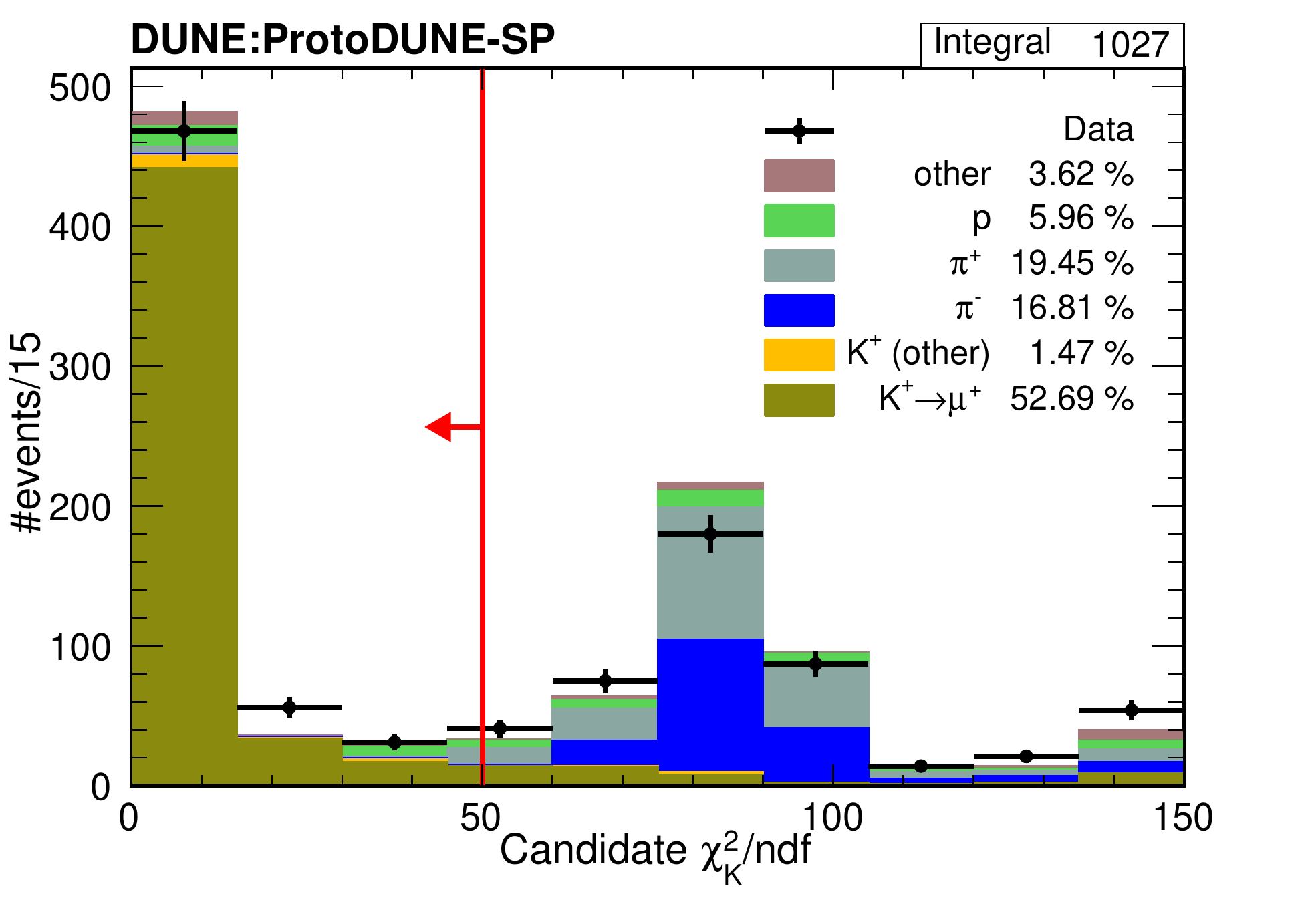}
\caption{Cut on the reduced $\chi^{2}$ distribution under kaon hypothesis to isolate the kaons from the remaining backgrounds. This sample is obtained after applying all but the last cut of the selection. Only statistical uncertainties of the data sample are shown, and the simulation is scaled to the data sample based on the number of candidates.
\label{fig:selection_result}}
\end{figure}

As it can also be observed, initially the kaon candidates represent only 1.36\% of the total candidates. After applying the daughter cuts, the purity increases up to 29.08\%. At this point, the dominant background is formed by inelastically interacting $\pi^{\pm}$. The use of geometric information in the $\cos{\theta_{\mathrm{can-dau}}}$ and the parent-daughter distance cuts increases the purity up to 52.69\%. Finally, the $\chi^{2}_{K}$ cut is applied to separate kaons from the remaining $\pi^{\pm}$. This cut is represented in Fig.~\ref{fig:selection_result}, where two populations are clearly distinguishable: one with low $\chi^{2}_{K}$, corresponding to stopping kaons and some proton contamination, and another one with high $\chi^{2}_{K}$, corresponding to the background. One can see that a very relaxed cut is enough to get a very pure sample of kaons. After applying this last cut, the remaining background is dominated by stopping protons, and it constitutes approximately a 3\% of the selected sample. 

\begin{table*}[htbp]
\begin{center}
\caption{Event selection summary for data and simulation as a function of the selection cuts. The evolution of the number of signal candidates and the respective efficiency and purity refer only to the simulation sample. No scaling applied. The mismatch in the number of signal events between the Purity and Efficiency columns is due to the aforementioned reconstruction issues caused by the electron diverters, which produced broken tracks: multiple reconstructed tracks in a single event can be associated to the same true object.}
\label{tab:selectionresult}  
\begin{tabular}{ l | c  c | c  c }
	\hline
	\hline
	\hline 
	\multirow{2}{*}{Selection Cut}        & \multicolumn{2}{c|}{Selected candidates}  & \multicolumn{2}{|c}{Signal candidates} \\ 
	    					     & ~~~Data~~~   & ~~~Simulation~~~ & Selected (Purity \%) & True (Efficiency \%) \\
	\hline
	True Candidates             & --     & --         & --           & 37792 (100.00)        \\
	Candidate Reconstructed     & 310276 & 313423     & 4276 (1.36)  & 4090 (10.82)        \\ 
	Fiducial Volume             & 192812 & 200690     & 3162 (1.58)  & 3009 (7.96)        \\
	Daughter Track-Like         & 132010 & 143409     & 2574 (1.79)  & 2454 (6.49)        \\
	Daughter $\chi^{2}_{\mu}$   & 21813  & 20781      & 1176 (5.66)  & 1170 (3.10)        \\
	Daughter Momentum by Range  & 2423   & 2280       & 663 (29.08)  & 663 (1.75)       \\
	Candidate-Daughter Angle    & 1335   & 1292       & 569 (44.04)  & 569 (1.51)       \\
	Candidate-Daughter Distance & 1027   & 1023       & 539 (52.69)  & 539 (1.42)       \\
	Candidate $\chi^{2}_{K}$     & 522    & 526       & 484 (92.02)  & 484 (1.28)        \\
	\hline
	\hline
	\hline 
\end{tabular}
\end{center}
\end{table*}

\section{ENERGY LOSS EVALUATION}
\label{sec:fit}
\noindent Proton decay searches at DUNE ultimately rely on the capabilities of the detector to identify kaons and to confidently differentiate them from other particles, especially from protons, which have the most similar signatures. This can be achieved by studying the energy loss profile along the trajectory of stopping particles, particularly in the Bragg peak region. Since future proton decay searches at DUNE will be blinded, it is fundamental that the simulation accurately reproduces this process. Here, we present the methodology developed to characterize the energy loss of kaons in LAr in detail. 

\begin{figure}[htbp]
\centering
\includegraphics[width=0.45\textwidth]{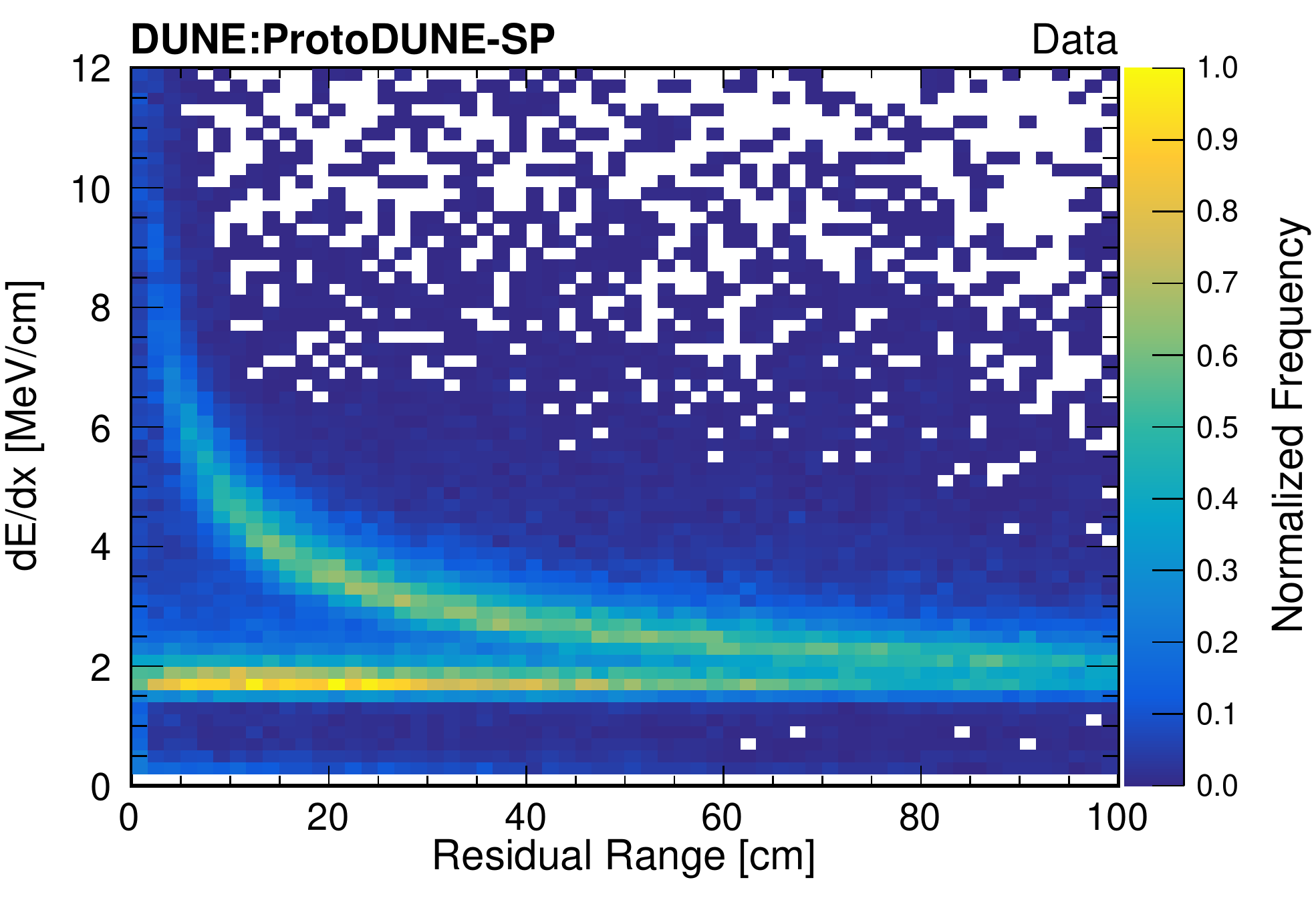}
\includegraphics[width=0.45\textwidth]{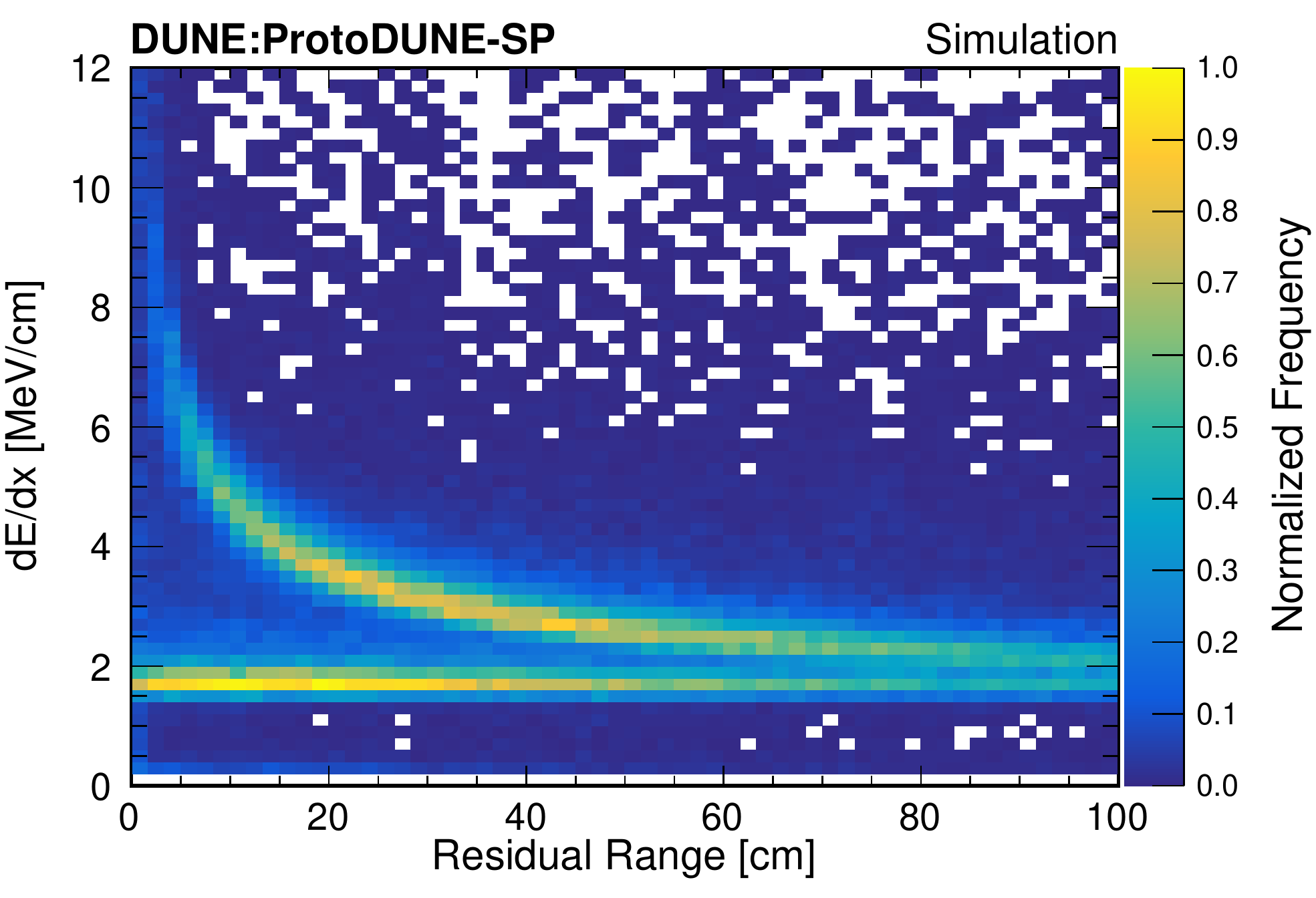}
\caption{$\mathrm{d}E/\mathrm{d}x$ distribution as a function of the residual range for the sample presented in Fig.~\ref{fig:selection_result} for data (top) and simulation (bottom).
\label{fig:dedx_2d}}
\end{figure}

In a LArTPC, the particle identification is usually based on the $\mathrm{d}E/\mathrm{d}x$ profile as a function of the residual range, as shown in Refs.~\cite{bib:dedx_resrange,bib:argoneut_dedx,bib:resrange_2}. A particle track consists of a collection of energy deposits (hits) in the three-dimensional space, and for each hit the residual range is defined as the distance from that particular hit to the endpoint of the track. Thus, in the case of stopping particles, hits closer to the track endpoint (low residual range) have larger energy depositions than hits farther away (high residual range). This effect can be seen in Fig.~\ref{fig:dedx_2d} for both data (top) and simulation (bottom), where this two-dimensional plot is presented for the sample shown in Fig.~\ref{fig:selection_result}. Two populations are clearly distinguishable for data and simulation. The population in the bottom of the plot, which shows a flat $\mathrm{d}E/\mathrm{d}x$ distribution along the residual range, corresponds to the interacting $\pi^{\pm}$ background. The other population, corresponding to the stopping kaons, shows an increase in the $\mathrm{d}E/\mathrm{d}x$ ---the Bragg peak--- as residual range decreases. Since the $\chi^{2}_{Part}$ is a $\mathrm{d}E/\mathrm{d}x$-based quantity, the last step of the selection presented above is not applied here to avoid any potential bias in the evaluation of kaons' $\mathrm{d}E/\mathrm{d}x$. 

\begin{figure}[htbp]
\centering
\includegraphics[width=0.45\textwidth]{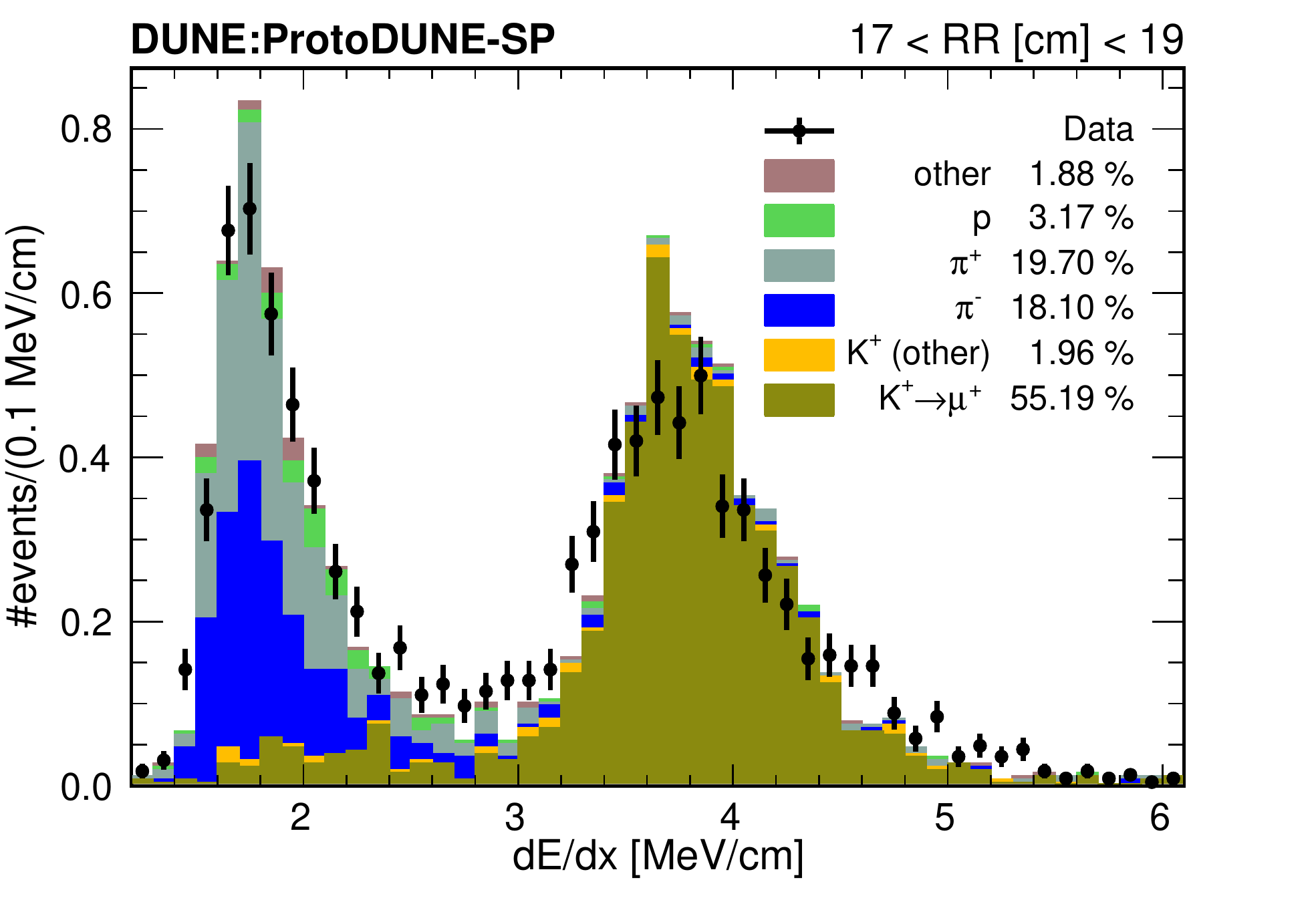}
\caption{$\mathrm{d}E/\mathrm{d}x$ distribution for the residual range slice between 17 and 19 cm. Each entry corresponds to a hit of a selected candidate. Data and simulation samples are normalized such that their integrated area is 1. Only data statistical uncertainties are shown. 
\label{fig:dedx_slice}}
\end{figure}

Consequently, in order to separate the contributions of kaons and background, we have developed a procedure to characterize the energy loss of kaons similar to the one developed by the ALICE collaboration to measure the hadron composition in charged jets from $pp$ collisions~\cite{bib:coherent_fit,bib:xianguo}, the so-called Coherent Fit. We divide the two-dimensional plot in residual range slices, generating one-dimensional $\mathrm{d}E/\mathrm{d}x$ distributions as the one presented in Fig.~\ref{fig:dedx_slice}. As the region of interest for this study is the Bragg peak, we limit our focus to residual ranges between 1 to 61~cm. We use 2~cm-long slices to find a compromise between resolution and statistics. 

The slices are modeled by a sum of Landau-Gaussian convolutions $\mathcal{LG}$. We choose this approach so that the Landau describes the energy loss and the Gaussian accounts for detector resolution and smearing produced by the residual range slicing. Furthermore, as the $\mathrm{d}E/\mathrm{d}x$ presents different shapes along its range (being more Landau-like for low $\mathrm{d}E/\mathrm{d}x$ distributions and more Gaussian for high $\mathrm{d}E/\mathrm{d}x$ distributions~\cite{bib:pdsp_tlfa}), the convolution can accommodate both shapes. We describe the Landau distribution following Ref.~\cite{bib:landau}, where three parameters are used: the normalization $N$, a `location' parameter $\mu$ and a `width' parameter $\sigma_{L}$, in such a way that it transforms as $\mathcal{L}(t)\rightarrow\mathcal{L}(\frac{x-\mu}{\sigma_{L}})$. The Landau distribution is convolved with a Gaussian distribution centered in $\mu$ with a standard deviation $\sigma_{G}$. Thus, the $k$-th Landau-Gaussian convolution $\mathcal{LG}$ of the sum is described by four parameters, $\vec{a}_{k}=(N_{k},\mu_{k},\sigma_{Lk},\sigma_{Gk})$. It is worth noticing that the parameter $\mu$ does not exactly correspond to the most probable value (MPV) of the Landau distribution, but this can be recovered afterwards by obtaining the maximum value. In the same way, $\sigma_{L}$ and $\sigma_{G}$ are related to the spread of the $\mathrm{d}E/\mathrm{d}x$ distribution on each slice, being the former related with the energy loss fluctuations and the later with the detector resolution and residual range smearing. 

We choose to describe each slice as a sum of three $\mathcal{LG}$ functions. One is used to describe the signal peak $S$, which is the one at high $\mathrm{d}E/\mathrm{d}x$ in Fig.~\ref{fig:dedx_slice} (mostly populated by $K^{+}$ hits). The other two are used to describe the background, which constitutes the peak at low $\mathrm{d}E/\mathrm{d}x$ (mostly formed by $\pi^{\pm}$) and the intermediate region between peaks, referred to as $B1$ and $B2$ respectively. Then, the number of counts in the residual range slice $i$ and $\mathrm{d}E/\mathrm{d}x$ bin $j$ is modeled as 

\begin{equation}
	M\left(RR_{i}, \left. \frac{\mathrm{d}E}{\mathrm{d}x}\right\rvert_{j}\right) = \sum_{k}^{S,B1,B2}\mathcal{LG}\left(\left. \frac{\mathrm{d}E}{\mathrm{d}x}\right\rvert_{j};\vec{a}_{k}\right) \mathrm{.}
\end{equation}

\noindent Assuming that these parameters vary continuously and smoothly from one residual range slice to another, we derive empiric functions to describe $\vec{a}_{k}$ as functions of the residual range, namely $\vec{a}_{k}\equiv\vec{a}_{k}(RR,\vec{b}_{k})$, where $\vec{b}_{k}$ is a different set of parameters. Hence, we can describe the model as


\begin{equation}
	M\left(RR_{i},\left. \frac{\mathrm{d}E}{\mathrm{d}x}\right\rvert_{j}\right) = \sum_{k}^{S,B1,B2}\mathcal{LG}\left(RR_{i},\left. \frac{\mathrm{d}E}{\mathrm{d}x}\right\rvert_{j};\vec{b}_{k}\right) \mathrm{.}
\end{equation}

\noindent In this way, instead of having a model per residual range slice, we have a single model that describes all slices simultaneously and hence it relies on a single minimization procedure. By construction, this approach considers correlations between slices, maximizes the use of information and reduces the effect of the limited statistics. For the optimization procedure we consider a maximum likelihood method in which the function $l$ is maximized 

\begin{equation}
l = \sum_{i}\sum_{j}P\left[f\left(RR_{i},\frac{\mathrm{d}E}{\mathrm{d}x}_{j}\right);M\left(RR_{i},\frac{\mathrm{d}E}{\mathrm{d}x}_{j}\right)\right] \mathrm{,}
\end{equation}

\noindent where $f$ is the measured distribution and it is normalised in every slice such that

\begin{equation}
\sum_{j}f\left(RR_{i},\frac{\mathrm{d}E}{\mathrm{d}x}_{j}\right) = 1 \mathrm{,}
\end{equation}

\noindent and $P$ represents the Poisson probability density function. This optimization provides the set of parameters $\vec{b_{k}}$ making the distribution $M$ more likely to have generated the measured distribution $f$ slice by slice and bin by bin.

We find empirical functions to describe the parameters $\vec{a}_{k}$ as functions of the residual range by using the truth information from the simulation. For the signal contribution these functions are

\begin{equation}
\label{eq:ts_mpv_param}
	\mu_{S}(RR) = \alpha_{\mu}\left(\frac{\beta_{\mu} RR-1}{\beta_{\mu} RR+1}+\gamma_{\mu}\right)
\end{equation}

\begin{equation}
\label{eq:ts_lw_param}
	\sigma_{L,S}(RR) =  \frac{1}{RR}\left(\frac{\alpha_{L}}{RR}-1\right)+\beta_{L}
\end{equation}

\begin{equation}
\label{eq:ts_gw_param}
	\sigma_{G,S}(RR) = \frac{\alpha_{G}}{RR+1}+\beta_{G} \mathrm{,}
\end{equation}

\noindent where the different $\alpha$, $\beta$ and $\gamma$ represent the parameters that continuously describe $\mu_{S}$, $\sigma_{L,S}$ and $\sigma_{G,S}$. Normalization is left constant across the slices, based on the observation that the $K^{+}$ hit purity is approximately 55\% and constant between 0 and 200~cm of residual range. For the background contribution we use

\begin{equation}
\label{eq:tb_mpv_param}
	\mu_{B1}(RR) = \alpha_{B1}+\beta_{B1} RR
\end{equation}

\begin{equation}
\label{eq:wil_lw_param}
	\sigma_{L,B2}(RR) = \frac{\alpha_{B2}}{RR+1}+\beta_{B2} \mathrm{.}
\end{equation}

\noindent All the other parameters are kept constant along the slices. Hence, instead of describing the 30 residual range slices with 3 Landau-Gaussian convolutions depending on four different parameters each (total of 360 parameters), this model attempts to describe all the data considering only 17 parameters. 

The minimization procedure is done using MINUIT~\cite{bib:minuit}. In order to guarantee an easier convergence, the following steps are followed. A first estimation of the parameters is done using the truth information from the simulation on each slice, which allows to study separately kaons and background. Then, the simulation sample is fitted coherently using the previous values as the seed. Finally, the data sample is coherently fitted using the fit result of the simulation as a seed.

\section{EVALUATION OF UNCERTAINTIES}
\label{sec:syst}

\noindent While statistical uncertainties are calculated by the MINOS algorithm~\cite{bib:minos_1,bib:minos_2} during the fit procedure, the main sources of systematic uncertainties considered in this study are the calorimetric calibration, the space-charge effect, reconstruction effects related to the electron diverters, and the normalization of the proton background. Their impact is evaluated in the simulation sample using 1000 toy experiments, in which each source is varied within its uncertainty according to a Gaussian distribution. Observables derived from or dependent on these sources are modified accordingly prior to the event selection, resulting in 1000 different pseudo-experiments. Each set is then processed through the fitting procedure described above, providing the estimates of systematic uncertainty on both the $\mathrm{d}E/\mathrm{d}x$ characterization and the event selection.

As previously presented in Section \ref{sec:simreco}, the calorimetric response of ProtoDUNE-SP's TPC is affected by different physics processes and thus needs to be calibrated. Following the procedure developed by MicroBooNE~\cite{bib:argoneut_dedx,bib:MicroBooNE_cali}, the uncertainty of the calibration was estimated to be 2\%~\cite{bib:phd_diurba}, however a 3\% uncertainty is assumed to account for variations in the LAr purity over time. 

Even though the effect on the calorimetry is considered in the previous uncertainty, the SCE also affects the spatial reconstruction of the  events. The overall spatial calibration of the SCE has an uncertainty of 8\%, which is estimated using the spread in the mean distortion of cosmic rays at the surfaces of the detector over time. 

Malfunctioning electron diverters between APAs were grounded during ProtoDUNE-SP operation, resulting in charge loss. In consequence, some tracks spanning more than one APA are misreconstructed: the track is split in two in the region between the APAs, and the second half is commonly assigned as a daughter of the first half~\cite{bib:pdsp_firstresults,bib:pdsp_k_xs}. This effect was included in the simulation, but was overestimated, leading to an excess of shorter tracks. This difference in data and simulation is measured by quantifying the number of broken tracks using 1 GeV/c muons from the beam at different angles with respect to the APA plane. It is found that broken tracks in simulation need to be weighted by a factor of $0.50\pm0.06$.

Protons and kaons present a similar $\mathrm{d}E/\mathrm{d}x$ profile as a function of the residual range. Although the proton presence in the sample presented in Fig.~\ref{fig:selection_result} is small (5\%), a difference in the proton contamination in the data and simulation samples could potentially bias the $\mathrm{d}E/\mathrm{d}x$ characterization. To address this, we apply the event selection to beam protons with a single reconstructed daughter in 1, 2, and 3 GeV/c beam runs for data and simulation. We observe that the proton background in the simulation sample needs to be weighted by a factor of $2.3\pm1.0$ in order to match that of the data sample. 

Table~\ref{tab:errors} presents the effect of these individual contributions on the total uncertainty considered in this study. 

\begin{table}[htbp]
\begin{center}
\caption{Effect of different uncertainties on the $\mathrm{d}E/\mathrm{d}x$ MPV. The last column presents the effect of all the uncertainties propagated simultaneously, in such a way that those affecting other are applied first.}
\label{tab:errors}  
\begin{tabular}{ l | c }
	\hline
	\hline
	\hline 
	Source             & Relative uncertainty (\%) \\
	\hline
	Statistical        & 1.1 \\
	Calorimetry        & 3.4 \\
	SCE                & 0.4 \\
	Electron diverters & 0.1 \\
	$p$ background     & 0.1 \\
	\hline 
	Total correlated   & 3.8 \\
	\hline
	\hline
	\hline 
\end{tabular}
\end{center}
\end{table}

\section{RESULTS}
\label{sec:results}
\subsection{d\textit{E}/d\textit{x} characterization}
\begin{figure*}[htbp]
\centering
\includegraphics[width=0.45\textwidth]{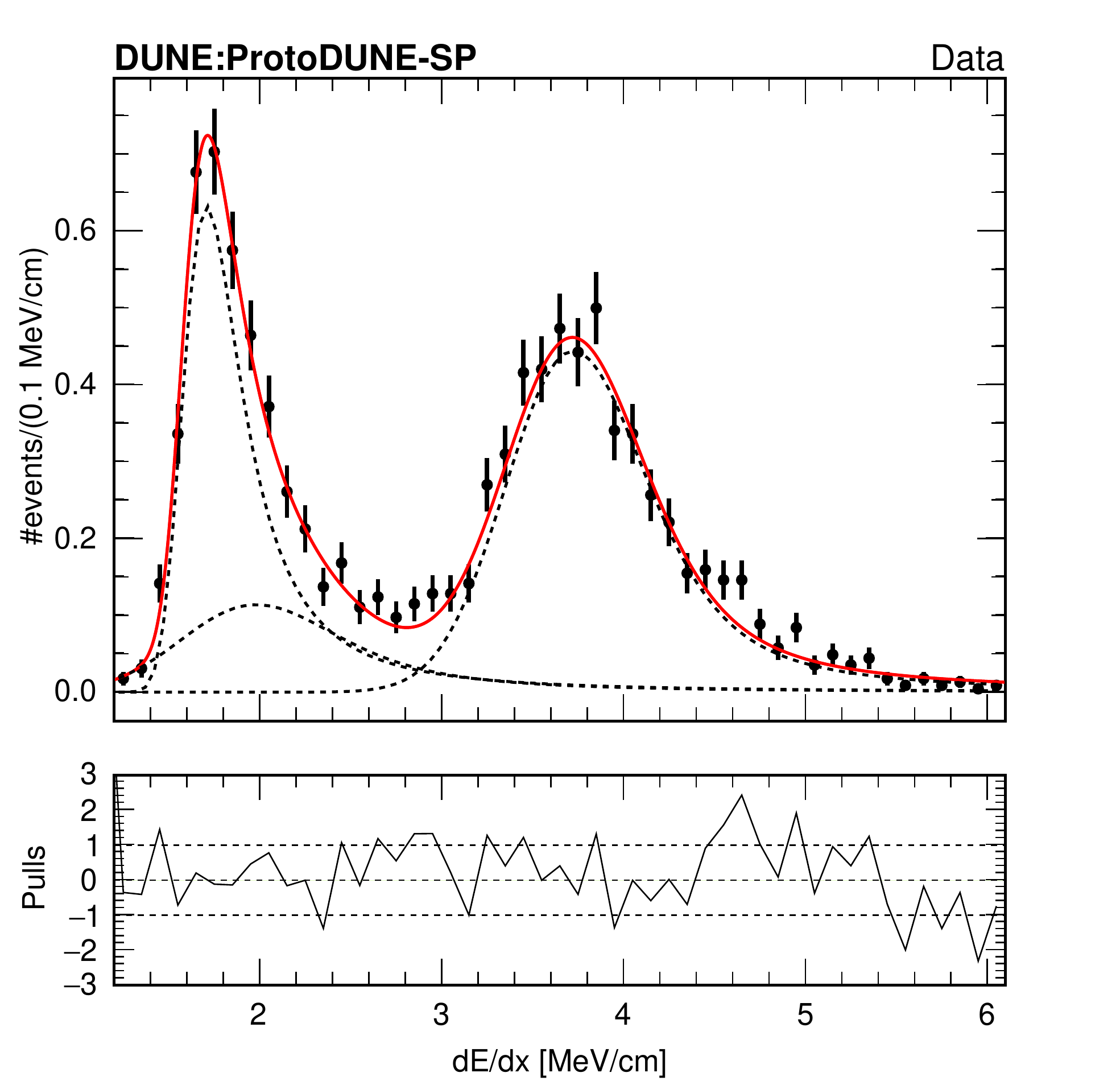}
\includegraphics[width=0.45\textwidth]{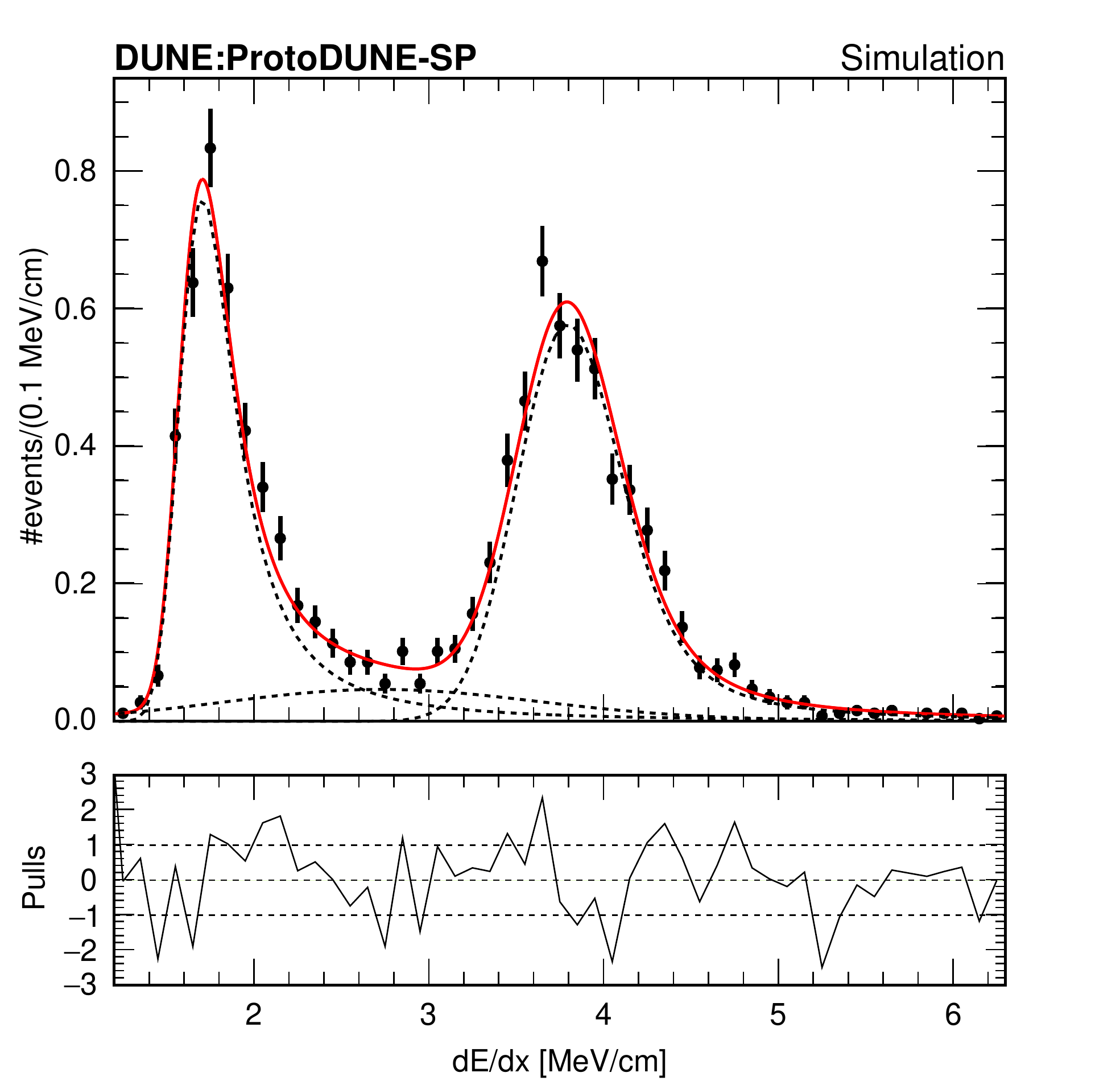}
\caption{$K^{+} \mathrm{d}E/\mathrm{d}x$ distribution for residual ranges between 17 and 19 cm for data (left) and simulation (right). The red line represents the complete model. The dashed lines represent the individual Landau-Gaussian convolution used: the one at higher $\mathrm{d}E/\mathrm{d}x$ describing the kaons' energy loss, and the other two and lower $\mathrm{d}E/\mathrm{d}x$ describing the energy loss of the backgrounds. The pulls are displayed in the panel below each plot.
\label{fig:fit_example}}
\end{figure*}

\begin{figure*}[htbp]
\centering
\includegraphics[width=0.45\textwidth]{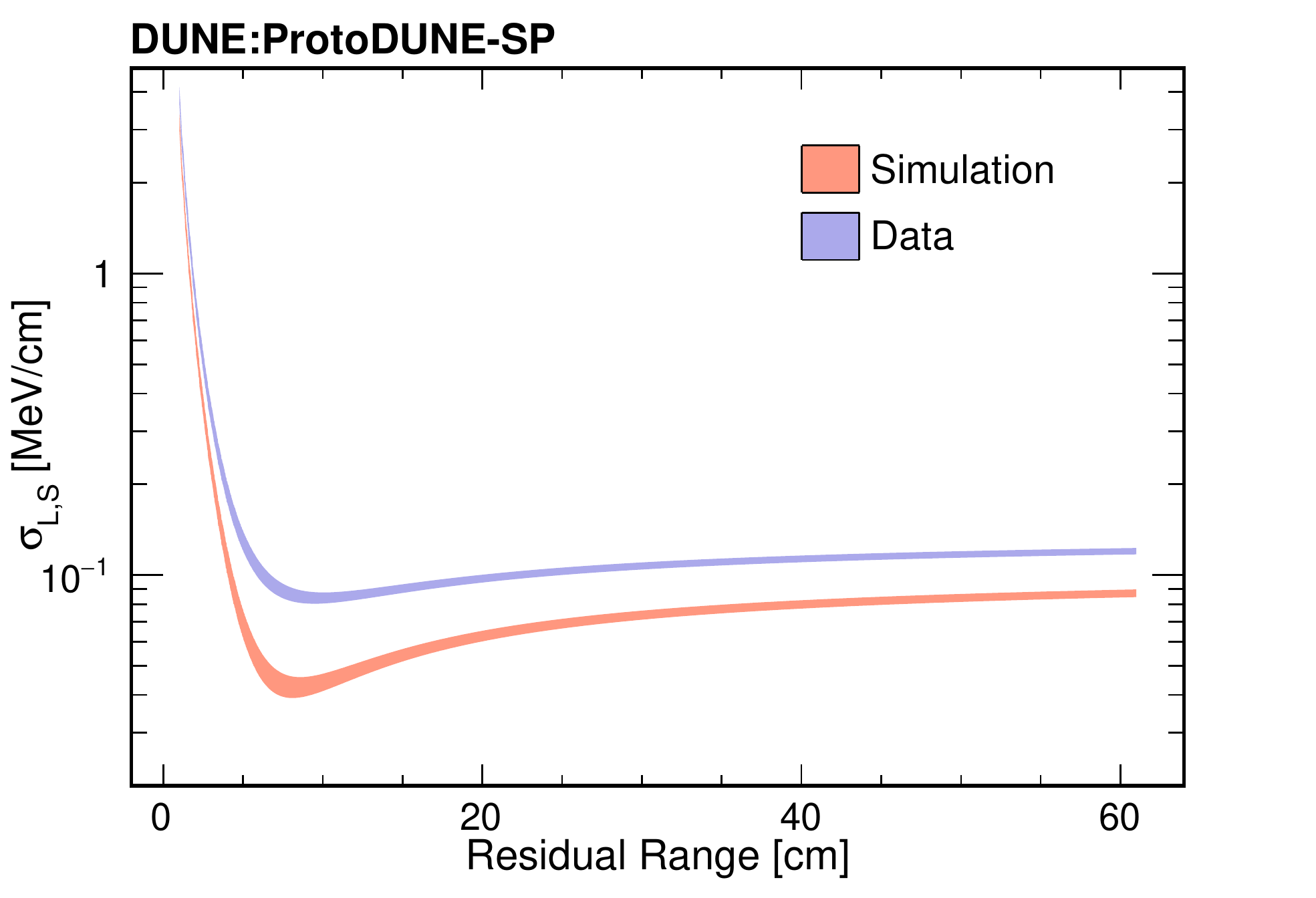}
\includegraphics[width=0.45\textwidth]{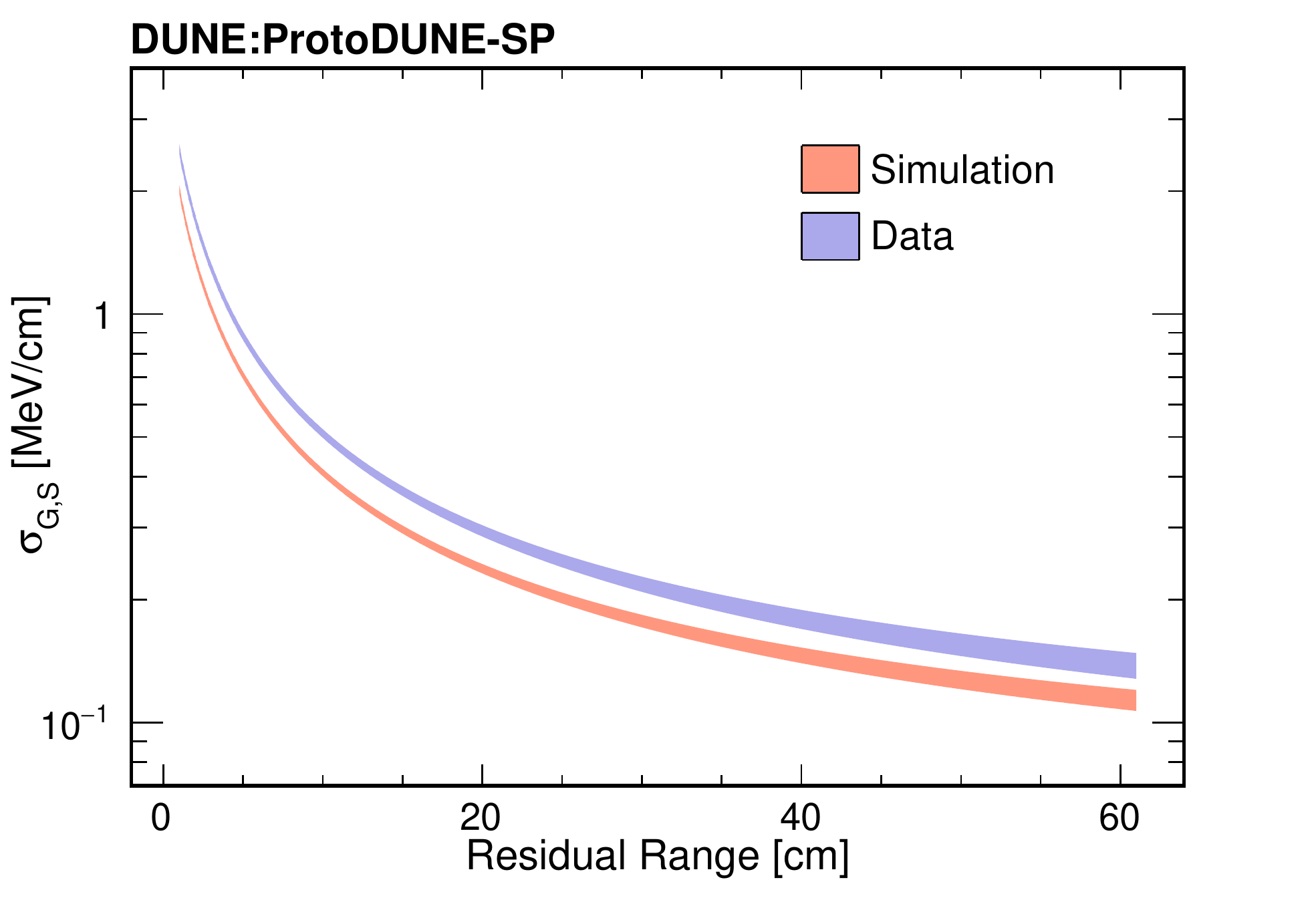}
\includegraphics[width=0.45\textwidth]{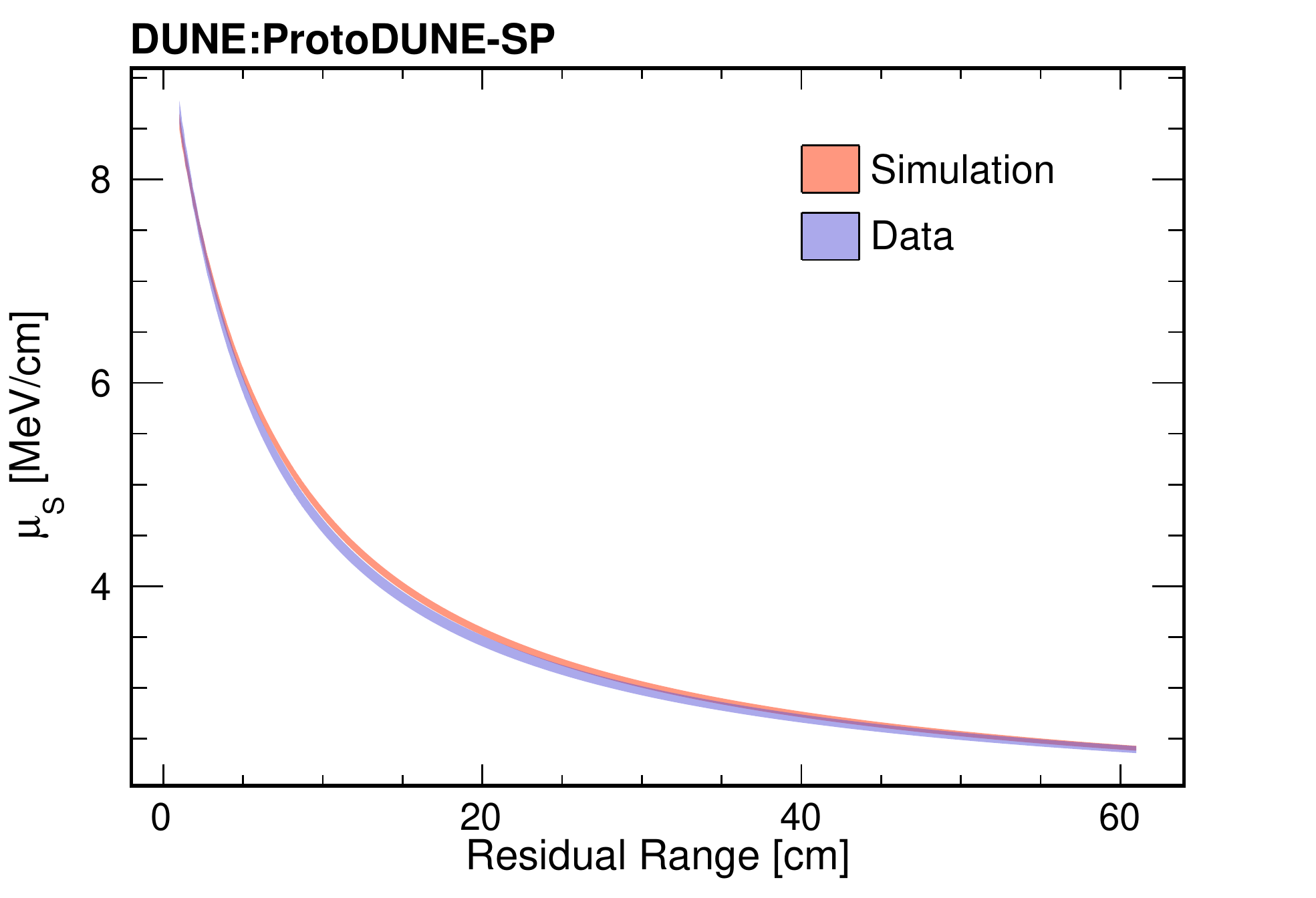}
\includegraphics[width=0.45\textwidth]{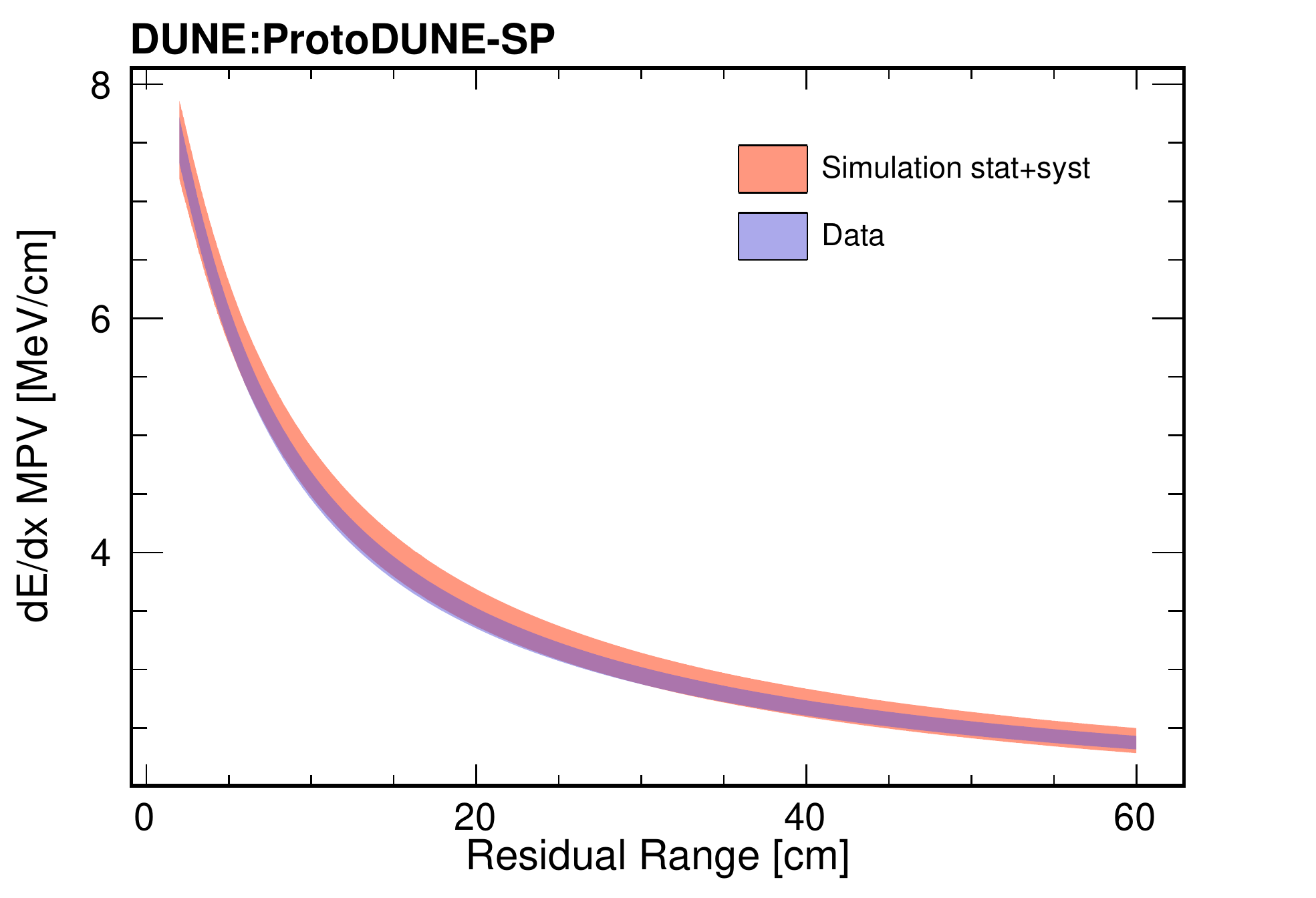}
\caption{$\sigma_{L,S}$ (top left), $\sigma_{G,S}$ (top right), $\mu_{S}$ (bottom left) as a function of residual range with statistical uncertainties only, and $K^{+} \mathrm{d}E/\mathrm{d}x$ most probable value (bottom right) as a function of residual range, with both statistical and systematic uncertainties.
\label{fig:mpv}}
\end{figure*}

\noindent The result of the fit procedure presented in Sec.\ref{sec:fit} provides a continuous representation of $K^{+}$ $\mathrm{d}E/\mathrm{d}x$ as a function of residual range. Fig.~\ref{fig:fit_example} shows an example of the obtained description for the residual range slice between 17 and 19 cm for data (left) and simulation (right). The distribution of the pulls of the fit is presented in the bottom panel of both plots. These are defined as the difference between the histogram and the value reported by the fit, divided by the uncertainty of the data points. The obtained parameters for the signal distribution are presented in Table~\ref{tab:fit_results}. The continuous representation of $\sigma_{L,S}$, $\sigma_{G,S}$ and $\mu_{S}$ are presented in Fig.~\ref{fig:mpv}-top-left, top-right and bottom-left, respectively, for data and simulation. While a similar behaviour is observed for $\mu_{S}$ in data and simulation, it can be noticed that the simulation presents smaller values for $\sigma_{L,S}$ and $\sigma_{G,S}$ than data. This is due to the energy resolution in the data sample being worse than in the simulation, and it is not surprising as it could be observed already in Fig.~\ref{fig:dedx_2d} and~\ref{fig:dedx_slice}. A kink can be observed in Fig.~\ref{fig:mpv}-top-left, that corresponds to the energy region where the $\mathrm{d}E/\mathrm{d}x$ distribution transitions from a Gaussian shape to a Landau-like shape. We recover the $\mathrm{d}E/\mathrm{d}x$ most probable value by finding the maximum of the Landau for every residual range. This is presented in Fig.~\ref{fig:mpv}-bottom-right, for data and simulation, where both are in good agreement within the estimated uncertainties.

\begin{table}[htbp]
\begin{center}
\caption{Fit results for the data sample.}
\label{tab:fit_results}  
\begin{tabular}{ c | c | c }
	\hline
	\hline
	\hline 
	$\vec{a}(RR,\vec{b}$)                  & $\vec{b}$         & Best fit $\pm$ stat $\pm$ syst \\
	\hline
	\multirow{3}{*}{$\mu$ [MeV/cm]}        & $\alpha$ [MeV/cm] & 4.125  $\pm$ 0.070 $\pm$ 0.130 \\
	                                       & $\beta$           & 0.189  $\pm$ 0.006 $\pm$ 0.003 \\
	                                       & $\gamma$          & -1.420 $\pm$ 0.006 $\pm$ 0.003 \\
	\hline
	\multirow{2}{*}{$\sigma_{L}$ [MeV/cm]} & $\alpha$ [MeV cm] & 4.891  $\pm$ 0.200 $\pm$ 0.100 \\
	                                       & $\beta$  [MeV/cm] & 0.135  $\pm$ 0.003 $\pm$ 0.002 \\
	\hline
	\multirow{2}{*}{$\sigma_{G}$ [MeV/cm]} & $\alpha$ [MeV]    & 4.979  $\pm$ 0.160 $\pm$ 0.010 \\
	                                       & $\beta$  [MeV/cm] & 0.058  $\pm$ 0.009 $\pm$ 0.005 \\                          
	\hline
	\hline
	\hline 
\end{tabular}
\end{center}
\end{table}

\subsection{Identification of low-energy kaons}
\begin{figure}[htbp]
\centering
\includegraphics[width=0.45\textwidth]{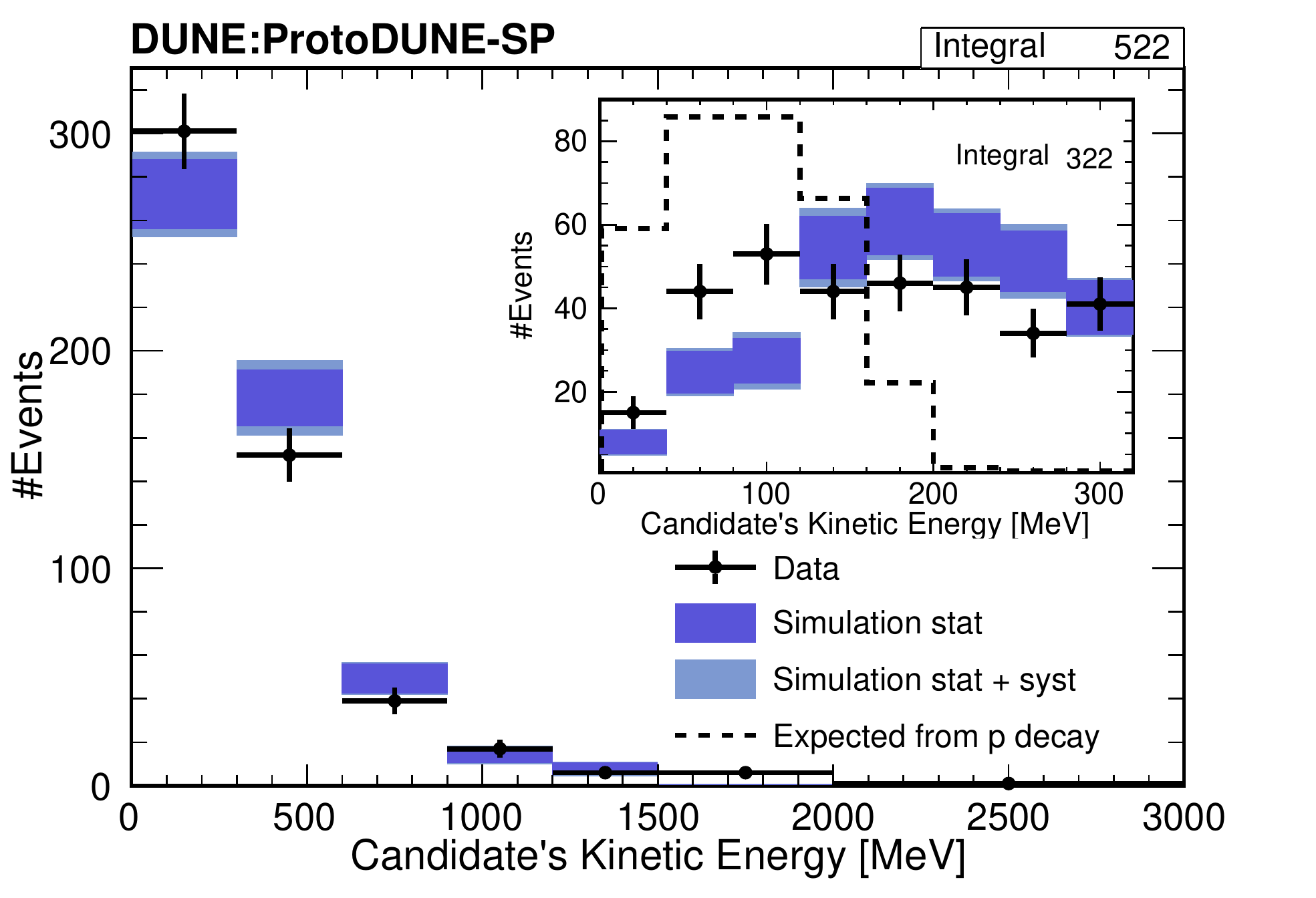}
\includegraphics[width=0.45\textwidth]{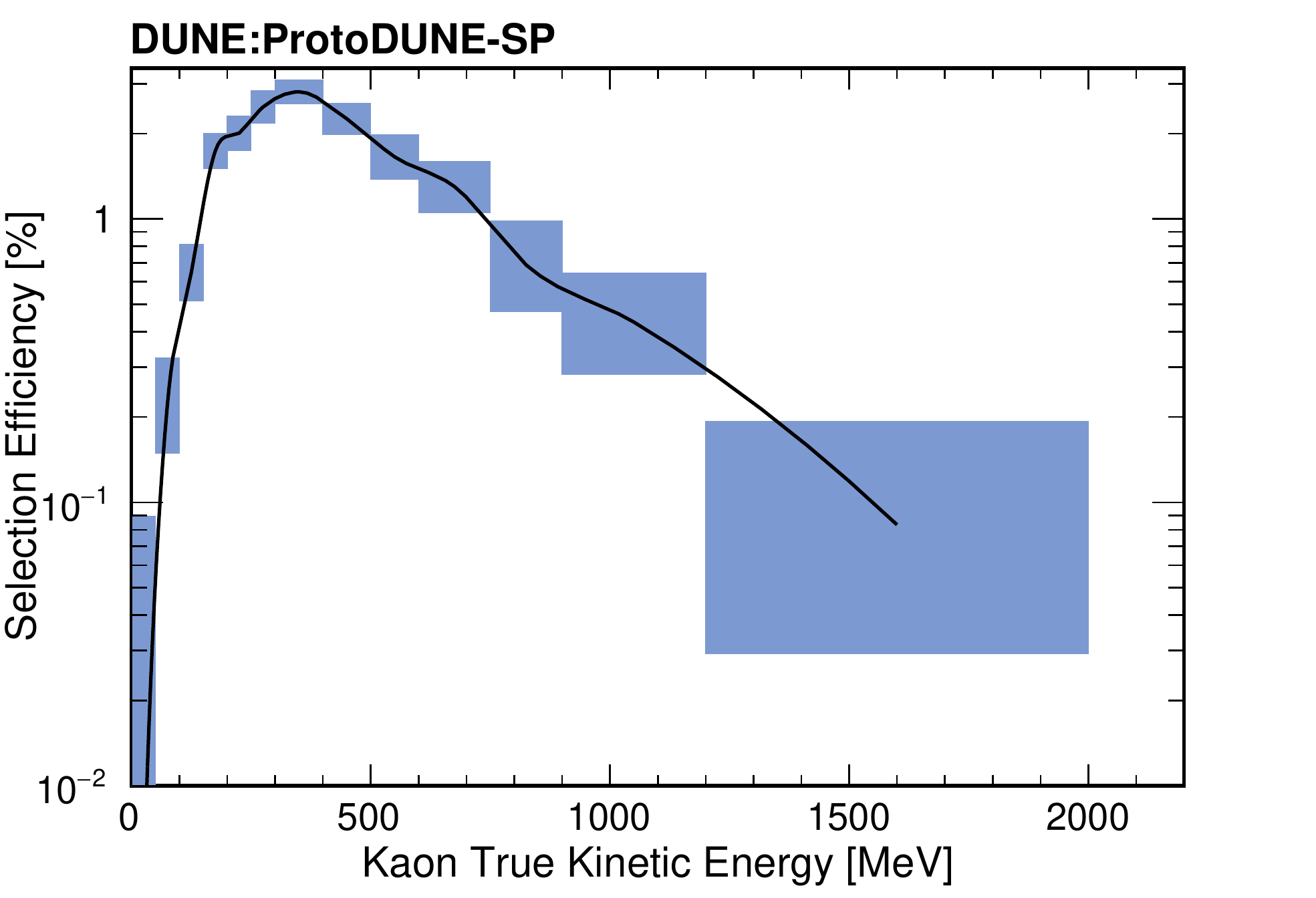}
\caption{Top: initial kinetic energy of the selected kaon sample. A detailed view of the energy range of interest for proton decay searches in DUNE is presented in the top right panel. It can be seen that 322 kaon candidates, spanning the expected kinetic energy resulting from proton decay, have been selected. Simulation sample and proton decay prediction are scaled to data based on the number of candidates. Bottom: selection efficiency as a function of the true kinetic energy of the kaon. Statistical and systematic uncertainties are shown, and the line is added as a guide to the eye. 
\label{fig:kaon_kinetic}}
\end{figure}

\noindent Fig.~\ref{fig:kaon_kinetic}-top shows the initial kinetic energy for the selected kaon sample. As these kaons are stopping, their kinetic energy can be estimated from their total deposited energy

\begin{equation}
	E_{K}=\sum_{i}^{N_{hits}}\left(\frac{\mathrm{d}E}{\mathrm{d}x}\right)_{i}\cdot \mathrm{d}x_{i} \mathrm{.}
\end{equation}	

\noindent An overall good agreement between data and simulation can be observed. A detail of the $K^{+}$ energy spectrum in the [0,320] MeV energy range is shown in the top right panel of the same plot, along with the energy distribution of $K^{+}$ resulting from proton decay in DUNE via $p \rightarrow K^{+}\bar{\nu}$. As it can be observed, the selected data sample covers the expected energy range for this hypothetical phenomenon, proving the capabilities of the LArTPC technology to study the proton decay phenomenon.

Fig.~\ref{fig:kaon_kinetic}-bottom shows the selection efficiency as a function of the true kinetic energy of the kaon. Although the selection conditions are different from those of proton decay searches in the DUNE FD, making the absolute efficiency scale less relevant here, several features are still noteworthy. The selection efficiency peaks between 150 and 500 MeV, corresponding to the higher end of the energy spectrum of kaons produced in proton decays. This is not surprising, as kaons in this energy range produce long enough tracks to be reconstructed, typically decaying at rest and passing the geometric cuts of the selection. Above 500 MeV, the efficiency decreases because kaons are less likely to decay at rest and inelastic scattering processes become more relevant. Below 150 MeV/c, the efficiency drops due to the kaon tracks being too short for reliable reconstruction or not passing the geometric cuts. Finally, no kaons with true kinetic energy below 50 MeV were selected. These observations are similar to the ones presented in~\cite{bib:dune_tdr2}, where kaon selection efficiency is very low below 40 MeV, and increases with energy until saturates around 100~MeV. 

\begin{figure}[htbp]
\centering
\includegraphics[width=0.45\textwidth]{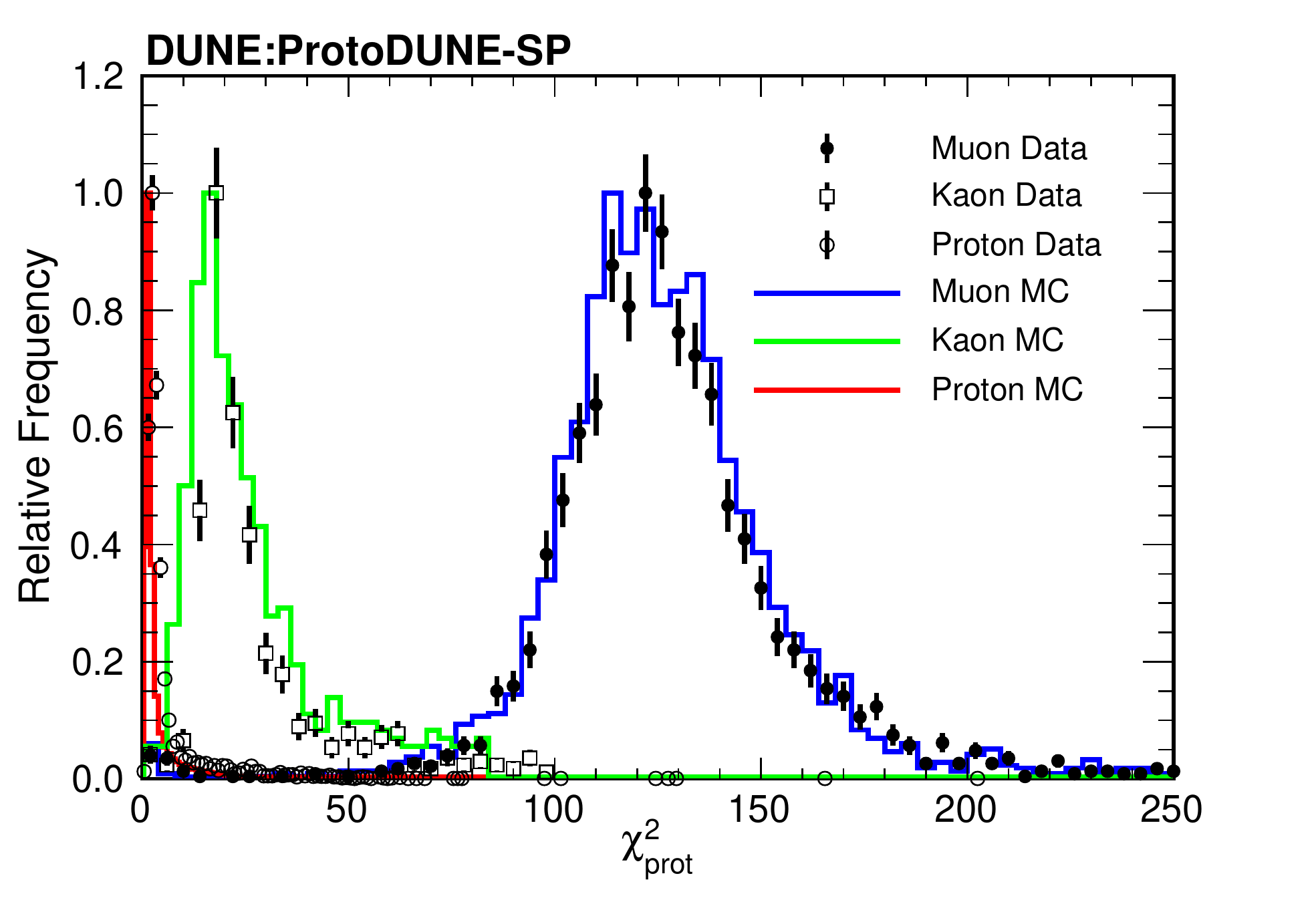}
\caption{Reduced $\chi^{2}$ under proton hypothesis for protons, kaons and muons. The samples have been normalized such that the maximum frequency is one. The samples of stopping protons and muons are the ones presented in \cite{bib:pdsp_performance}.
\label{fig:pid}}
\end{figure}

This sample of stopping kaons is compared in Fig.~\ref{fig:pid} with other samples of stopping protons and muons by means of the $\chi^{2}$ test assuming the proton hypothesis. As it can be noticed, three populations are clearly distinguishable in data and simulation, each one of them corresponding to a different particle type, demonstrating the PID capabilities of the LArTPC to confidently separate kaons from other particles.

\section{CONCLUSIONS}
\label{sec:conclusion}
\noindent This paper presents the first identification of low-energy kaons in LAr using ProtoDUNE-SP beam data. Analyzing the 6 and 7 GeV/c momenta beam runs taken during 2018, a selection of secondary stopping kaons produced in hadronic reactions has been presented. As a result, a population of 522 stopping kaons with a purity of 92\% has been obtained. This selection has been developed without relying on the Photon Detection System (PDS), highlighting the excellent performance of the charge readout of the detector. In DUNE FD, where the PDS will not be as saturated as in ProtoDUNE-SP, including its information is expected to further improve this identification by enhancing the energy resolution and by detecting the delayed flash of the Michel electron. 

The selected events span the energy range of interest for proton decay searches via $p \rightarrow K^{+}\bar{\nu}$. Although the detector conditions differ from those of DUNE’s far detector, these studies provide insight into the kaon detection efficiency. In particular, no kaons with energies below 50 MeV were selected, highlighting the need for dedicated approaches to identify these events, which account for 25\% of the kaons resulting from proton decays. Finally, this sample of stopping kaons has been compared with other samples of stopping protons and muons, and it has been shown that they can be distinguished and identified by means of a $\chi^{2}$ test of the energy loss profile, demonstrating the particle identification capabilities of the LArTPC technology.

In addition to this, the first detailed study of the kaon $\mathrm{d}E/\mathrm{d}x$ profile as a function of the residual range in LAr has been presented. The Coherent Fit approach has been used to evaluate the most probable value of kaons' $\mathrm{d}E/\mathrm{d}x$, which was found to be in good agreement with the simulation. However, some differences have been observed in the energy resolution between data and simulation, highlighting the need for further studies to achieve a more complete understanding of calorimetric measurements.

These results are of utmost importance, as they demonstrate the potential of the future DUNE project to address proton decay searches. 

\section{ACKNOWLEDGEMENTS}

%
%
\noindent The ProtoDUNE-SP detector was constructed and operated on the CERN Neutrino Platform.
We gratefully acknowledge the support of the CERN management, and the
CERN EP, BE, TE, EN and IT Departments for NP04/Proto\-DUNE-SP.
%
%
This document was prepared by DUNE collaboration using the resources of the Fermi National Accelerator Laboratory (Fermilab), a U.S. Department of Energy, Office of Science, Office of High Energy Physics HEP User Facility. Fermilab is managed by Fermi Forward Discovery Group, LLC, acting under Contract No. 89243024CSC000002.
%
%
This work was supported by
CNPq,
FAPERJ,
FAPEG and 
FAPESP,                         Brazil;
CFI, 
IPP and 
NSERC,                          Canada;
CERN;
ANID-FONDECYT,                  Chile;
M\v{S}MT,                       Czech Republic;
ERDF, FSE+,
Horizon Europe, 
MSCA and NextGenerationEU,      European Union;
CNRS/IN2P3 and
CEA,                            France;
PRISMA+,                        Germany;
INFN,                           Italy;
FCT,                            Portugal;
CERN-RO/CDI,                        Romania;
NRF,                            South Korea;
Generalitat Valenciana, 
Junta de Andaluc\'i­a-FEDER, 
MICINN, and 
Xunta de Galicia,               Spain;
SERI and 
SNSF,                           Switzerland;
T\"UB\.ITAK,                    Turkey;
The Royal Society and 
UKRI/STFC,                      United Kingdom;
DOE and 
NSF,                            United States of America.
%
%
This research used resources of the 
National Energy Research Scientific Computing Center (NERSC), 
a U.S. Department of Energy Office of Science User Facility 
operated under Contract No. DE-AC02-05CH11231.
%

\bibliography{./bibliography}

\end{document}